\documentclass{aastex61}

\received{December 2017}
\revised{February 2018}
\accepted{March 2018}
\submitjournal{ApJ}

\usepackage{longtable}
\usepackage{amssymb}
\usepackage{color}
\usepackage{wrapfig}
\usepackage{amsmath}
\usepackage{longtable}
\usepackage[version=3]{mhchem}
\usepackage[normalem]{ulem}
\usepackage{booktabs}

\newcommand{\feh}{$\mbox{[Fe/H]}$}

\begin{document}

\shorttitle{Chemical abundances of stars in Tucana II} 

\title{Chemical abundances of new member stars in the \\Tucana II dwarf galaxy\footnote{This paper includes data gathered with the 6.5\,m Magellan Telescopes located at Las Campanas Observatory, Chile. }}

\correspondingauthor{Anirudh Chiti}
\email{achiti@mit.edu}

\author{Anirudh Chiti}
\affiliation{Department of Physics and Kavli Institute for Astrophysics and  Space Research,  Massachusetts Institute of Technology, Cambridge, MA 02139, USA}

\author{Anna Frebel}
\affiliation{Department of Physics and Kavli Institute for Astrophysics and  Space Research,  Massachusetts Institute of Technology, Cambridge, MA 02139, USA}

\author{Alexander P. Ji}
\affiliation{Observatories of the Carnegie Institution of Washington, 813 Santa Barbara St., Pasadena, CA 91101, USA}
\affiliation{Hubble Fellow}

\author{Helmut Jerjen} 
\affiliation{Research School of Astronomy and Astrophysics, Australian National University, Canberra, ACT 2611, Australia}

\author{Dongwon Kim}
\affiliation{Astronomy Department, University of California, Berkeley, CA, USA}

\author{John E. Norris}
\affiliation{Research School of Astronomy and Astrophysics, Australian National University, Canberra, ACT 2611, Australia}

\begin{abstract}

We present chemical abundance measurements for seven stars with metallicities ranging from [Fe/H] = $-$3.3 to [Fe/H] = $-$2.4 in the Tucana II ultra-faint dwarf galaxy (UFD), based on high-resolution spectra obtained with the MIKE spectrograph on the 6.5\,m Magellan-Clay Telescope.
For three stars, we present detailed chemical abundances for the first time.
Of those, two stars are newly discovered members of Tucana II and were selected as probable members from deep narrow band photometry of the Tucana II UFD taken with the SkyMapper telescope. 
This result demonstrates the potential for photometrically identifying members of dwarf galaxy systems based on chemical composition. 
One new star was selected from the membership catalog of \citet{wmo+16}.
The other four stars in our sample have been re-analyzed, following additional observations.
Overall, six stars have chemical abundances that are characteristic of the UFD stellar population.
The seventh star shows chemical abundances that are discrepant from the other Tucana II members and an atypical, higher strontium abundance than what is expected for typical UFD stars.
While unlikely, its strontium abundance raises the possibility that it may be a foreground metal-poor halo star with the same systemic velocity as Tucana II.
If we were to exclude this star, Tucana II would satisfy the criteria to be a surviving first galaxy.
Otherwise, this star implies that Tucana II has likely experienced somewhat extended chemical evolution.

\end{abstract}

\section{Introduction} 
\label{sec:introduction}

The elements in the atmospheres of metal-poor stars allow us to study the chemical composition of the early universe.
The elements in stellar atmospheres reflect the composition of a star's formative gas cloud.
Thus, a low surface metal abundance of a metal-poor star indicates its natal gas cloud must have undergone relatively few cycles of chemical enrichments (e.g., from supernovae).
This lack of enrichment implies that metal-poor stars generally formed earlier than typical solar-metallicity stars, and that metal-poor stars can be used to probe the composition of the early universe in which they formed.

The iron abundance is typically used as a proxy for the overall metal context (or ``metallicity") of a star and metal-poor stars are defined to have an iron abundance of $\feh\,\,\le -1\,\text{dex}$, where [Fe/H] = $\log_{10}(N_{\text{Fe}}/N_{\text{H}})_{\star} - \log_{10}(N_{\text{Fe}}/N_\text{H})_\sun$ \citep{fn+15}.
Of particular interest are the most metal-poor stars, such as very metal-poor stars (VMP; $\feh\,\le-2.0$) and extremely metal-poor stars (EMP; $\feh\,\le-3.0$).
The abundance of various elements (i.e., carbon, neutron-capture elements) as a function of overall [Fe/H] for VMP and EMP stars sheds light on the nature of the chemical evolution of the early universe \citep{swp+96, bc+05, rpt+14, pfb+14}.
Stars with $\feh\,\,\le-4.0$ can be used to constrain the yields and properties of the very first supernovae \citep[e.g.,][]{hw+10} and by extension, properties of the first stars \citep{byh+09}.
Metal-poor stars have also been used to trace old substructure in the Milky Way (e.g., \citealt{smy+17}), and to address a number of questions related to galaxy formation and cosmology \citep{ss+82, fb+02,fcn+07,fn+13,kbb+13, fn+15}. 

The simpler formation history of dwarf galaxies makes them an ideal laboratory to use metal-poor stars for studying topics such as chemical evolution, star formation history, and stellar populations \citep{tht+09}.
Furthermore, faint dwarf galaxies are thought to be the surviving analogs of the ancient galaxies that were accreted to form the Milky Way halo \citep{fks+10, b+13}, and are also themselves older and more metal-poor than some components of the Milky Way such as the disk \citep{jg+07, kcg+13}.
Thus, studying the metal-poor stars in these systems provides insights on the nature of the first galaxies and the origins of the of chemical signatures of the VMP and EMP stellar population in the halo \citep{sh+15}.

Ultra-faint dwarf galaxies (UFDs), in particular, are among the oldest ($\geq$10\,Gyr), most metal-poor (typically a mean $\feh < -2.0$), and dark-matter dominated (M/L$_V$ $\gtrsim$ 100) \citep[e.g.,][]{btg+14} dwarf galaxy systems. 
These characteristics make stars in UFDs especially promising targets to study the aforementioned questions.
Several surveys over the past decade have detected dozens of UFDs \citep{wbd+05, zbe+06, bze+07, wjw+07, w+10, bdb+15, dbr+15, kj+15, kjm+15, kbt+15, lmi+15, lmb+15, dba+16, hco+16, hco+18}, thus greatly increasing the prospect for studying the population of metal-poor stars in their environments. 

To investigate the detailed chemical composition of stars, it is necessary to obtain high-resolution spectra. 
Results already show the utility of detailed studies of the composition of metal-poor stars in UFDs.
For instance, 
the strong over-abundance of neutron-capture elements associated with the r-process in seven stars in the Reticulum II UFD has constrained the dominant astrophysical site of the r-process \citep{jfc+16}. 
However, only 59 stars have been observed with high-resolution spectroscopy in 14 UFD systems 
\citep{kwg+08, fek+09, fks+10, fsg+10, fsk+14, ngw+10, nyg+10, jfw+10, gnm+13, kfa+13, iaa+14, kr+14, rk+14, jfe+16, jfs+16b, jfs+16, hsm+17, kcs+17, vsm+17, dms+17} since the low stellar mass ($\lesssim 10^4 M_\odot$), distance ($d \gtrsim 30\,\text{kpc}$), and lack of giant branch stars in UFDs \citep[e.g.,][]{mdr+08} strictly limits the stars for which high-resolution spectroscopy can be performed with current technology.
Adding to the observational burden, medium-resolution spectroscopy is required to identify which stars in their field are members of these systems before high-resolution observations can be carried out.
All of these reasons make the time required to identify and observe member stars of UFDs a bottleneck to progress in the field.

In this paper, we present the chemical abundances of seven stars with [Fe/H] ranging from $-2.4$ to $-3.3$\,dex in the UFD Tucana II \citep{bdb+15, kbt+15} derived from high-resolution spectroscopy. 
Two stars are new members that were identified from photometry of the Tucana II dwarf galaxy obtained with the filter set on the SkyMapper telescope (Chiti et al. 2018, in prep). 
The discovery of these stars motivated studying Tucana II in more detail.
We also observed one new star that was previously confirmed as a member by \citet{wmo+16}.
To supplement the new observations, we decided to re-analyze the four stars with published measurements from \citet{jfe+16}, after collecting additional data to improve measurement precision. 
Since observations suggest that UFDs contain no members with [Fe/H] $>$ $-1$, selecting metal-poor stars from photometry is a potentially powerful way to identify significant numbers of UFD members for spectroscopic follow-up observations. 
This has the potential for bypassing the expensive medium-resolution spectroscopy step of the process, and thus accelerating the characterization of UFDs and other dwarf galaxies.

This paper is organized as follows.
We outline the target selection procedure and observations in Section~\ref{sec:observations}, discuss the abundance analysis in Section~\ref{sec:abundances}, present the chemical signatures of stars in Tucana II and implications in Sections~\ref{sec:signatures}, and conclude in Section~\ref{sec:conclusions}.

\begin{deluxetable*}{llllllll} 
\tablecolumns{8}
\tablecaption{\label{tab:obs} Observations}
\tablehead{   
  \colhead{Name} &
  \colhead{RA (h:m:s) (J2000)} & 
  \colhead{DEC (d:m:s) (J2000)} &
  \colhead{Slit size} &
  \colhead{$g$ (mag)} &
  \colhead{$t_\text{exp}$ (min)} &
  \colhead{S/N$\tablenotemark{a}$} & 
  \colhead{$v_\text{helio}$ (km/s)} 
}
\startdata
TucII-006 & 22:51:43.06 & $-58$:32:33.7 & 1\farcs0 & 18.78 & 206$\tablenotemark{b}$ & 15, 30 & $-126.1$\\ 
TucII-011 & 22:51:50.28 & $-58$:37:40.2 & 1\farcs0 & 18.27 & 314$\tablenotemark{b}$ & 15, 30 & $-124.6$\\ 
TucII-033 & 22:51:08.32 & $-58$:33:08.1 & 1\farcs0 & 18.68 & 155$\tablenotemark{b}$ & 17, 32& $-126.9$\\ 
TucII-052 & 22:50:51.63 & $-58$:34:32.5 & 1\farcs0 & 18.83 & 155$\tablenotemark{b}$ & 17, 35& $-119.9$\\ 
TucII-078 & 22:50:41.07 & $-58$:31:08.3 & 1\farcs0 & 18.62 & 215 & 15, 30 & $-123.8$\\ 
TucII-203 & 22:50:08.87 & $-58$:29:59.1 & 1\farcs0 & 18.81 & 275 & 16, 37 & $-126.1$\\ 
TucII-206 & 22:54:36.67 & $-58$:36:57.9 & 1\farcs0 & 18.81 & 385 & 15, 37 & $-122.9$ \\
\enddata
\tablenotetext{a}{S/N per pixel is listed for 4500\,\AA\,\,and 6500\,\AA}
\tablenotetext{b}{Combined exposure time from \citet{jfe+16} and this work}
\end{deluxetable*}

\section{Target Selection and Observations}
\label{sec:observations}

\begin{figure*}[!htbp]
\centering
\includegraphics[width =0.99\textwidth]{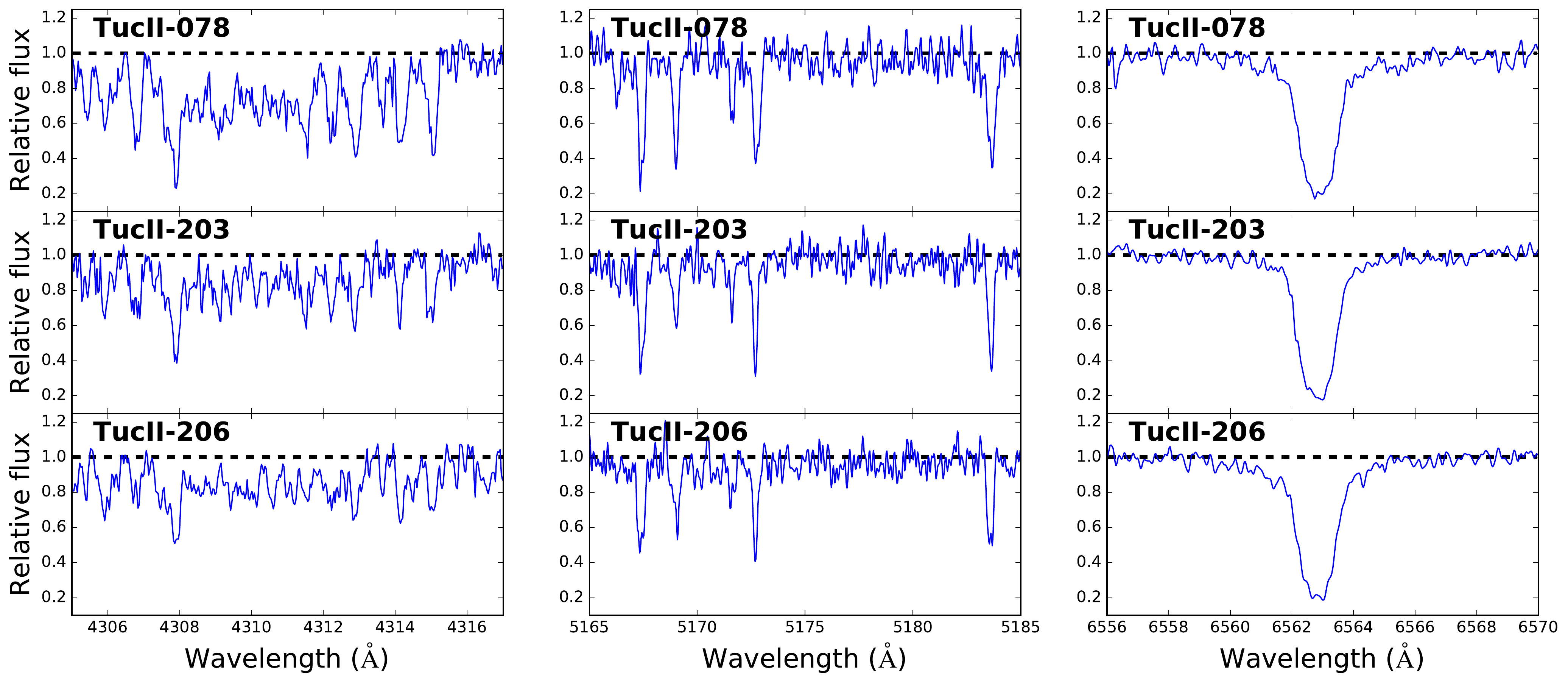}
\caption{Plots of the CH region (left), Mg b line region (center), and H$\alpha$ feature (right) for each of the Tucana\,II members with no prior high-resolution chemical abundance measurements available. 
TucII-078 was spectroscopically identified as member \citep{wmo+16}, while TucII-203 and TucII-206 were identified based on narrow band photometry.} 
\label{fig:samplespec}
\end{figure*}

\subsection{Members from \citet{wmo+16}}
\label{sec:obswalker}

\citet{jfe+16} observed TucII-006, TucII-011, TucII-033, and TucII-052 with the MIKE spectrograph (see Table~\ref{tab:obs}).
All four stars were selected from the membership catalog of \citet{wmo+16}.
They observed each star between 100\,mins to 4.42\,hrs in August 2016 with the MIKE spectrograph on the Magellan-Clay telescope.
For the stars with the shortest exposure times (TucII-033 and TucII-052), this precluded the measurement of several elements and led to large uncertainties in the measurement of the abundances of several other elements.  
Thus, we re-observed each star in \cite{jfe+16} for an additional 55\,mins to address the aforementioned deficiencies.
In addition to re-observing these stars, we observed an additional member (TucII-078) from \citet{wmo+16} that had not previously been observed with a high-resolution spectrograph.

\subsection{Members selected from SkyMapper photometry}
\label{sec:sky}

Through a P.I. program, we obtained SkyMapper photometry of Tucana II using the 1.3\,m telescope at Siding Spring Observatory. 
In an upcoming paper, we will fully discuss the implementation of the SkyMapper filter-set to determine photometric metallicities (Chiti et al. 2018, in prep) but we briefly discuss the method here.
The SkyMapper filter set includes a narrow-band $v$ filter that covers the prominent Ca II K line at 3933.7\AA\, \citep{bbs+11}.
Given the strength of this line, the preponderance or lack of metals sufficiently affects the line strength which changes the total flux through this filter.
Thus, a metal-poor star with a weak Ca II K line appears brighter in this filter than more metal-rich stars.
To quantify this effect, we generated a grid of flux-calibrated spectra using the Turbospectrum synthesis code \citep{ap+98,p+12}, the MARCS model atmospheres \citep{gee+08}, and a line list derived from the VALD database \citep{pkr+95, rpk+15}.
The stellar parameters of our grid covered the expected stellar parameters ($4000 < T_{\text{eff}}\,[\text{K}] < 5700; 1 < \log\,g < 3$) and metallicities ($-4.0 < \text{[Fe/H]} < -0.5$) of RGB stars in dwarf galaxies. 
We closely followed the methodology of \citet{bm+12} and \citet{cv+14} to generate a library of synthetic photometry through the SkyMapper $u$, $v$, $g$, and $i$ filters for spectra in this grid.

By relating our observed SkyMapper photometry in the $v$, $g$, and $i$ filters to the synthetic photometry from our grid, we selected a few metal-poor targets for spectroscopic test observations.
Two of these targets (TucII-203 and TucII-206) were confirmed as members of Tucana II since radial velocity measurements from their MIKE spectra were similar to the systemic velocity of Tucana II of  $-129.1\,\text{km/s}$ \citep{wmo+16}.

\subsection{High-resolution spectroscopy}

\begin{figure*}[!htbp]
\centering
\includegraphics[width =0.49\textwidth]{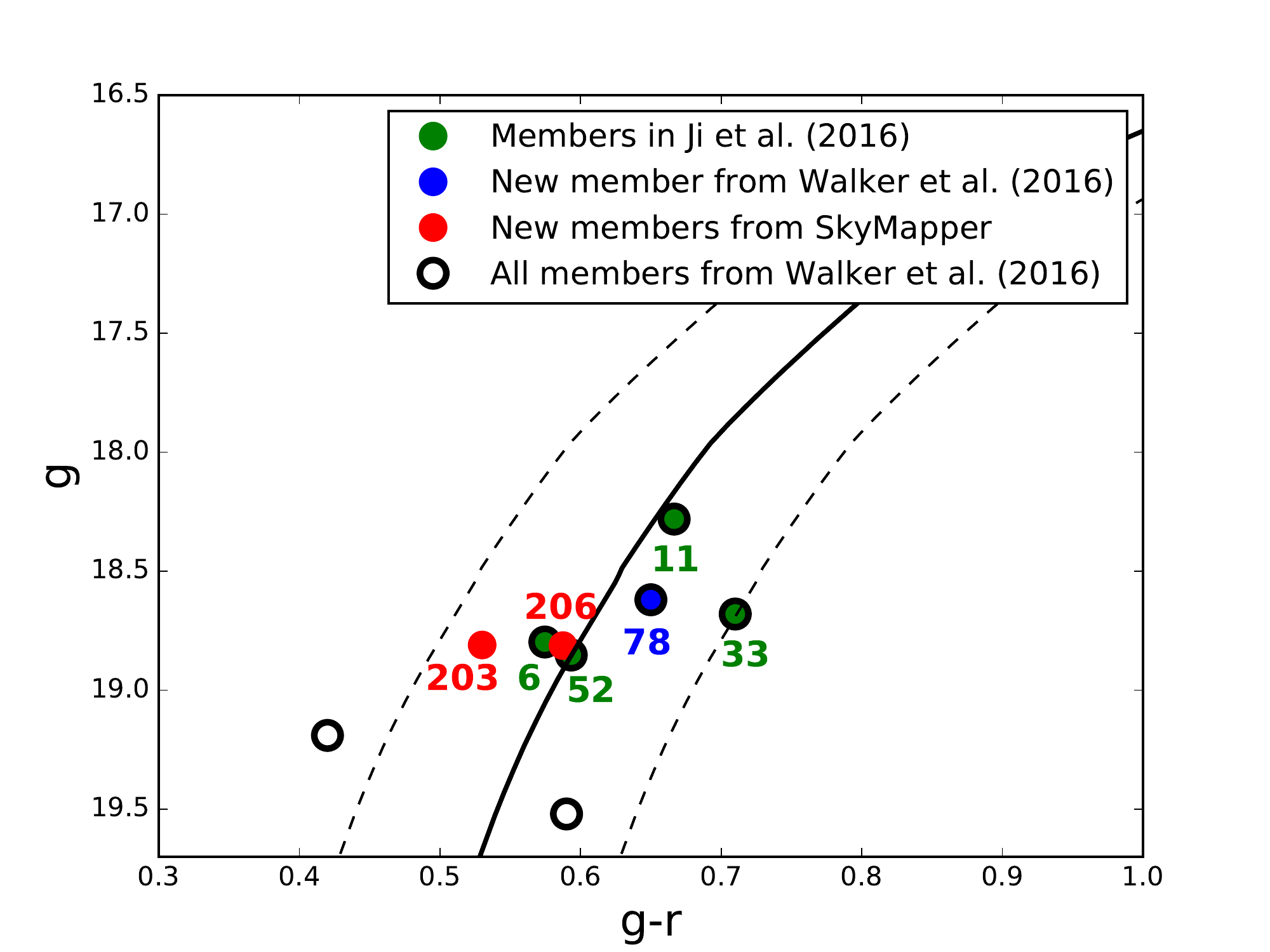}
\includegraphics[width =0.49\textwidth]{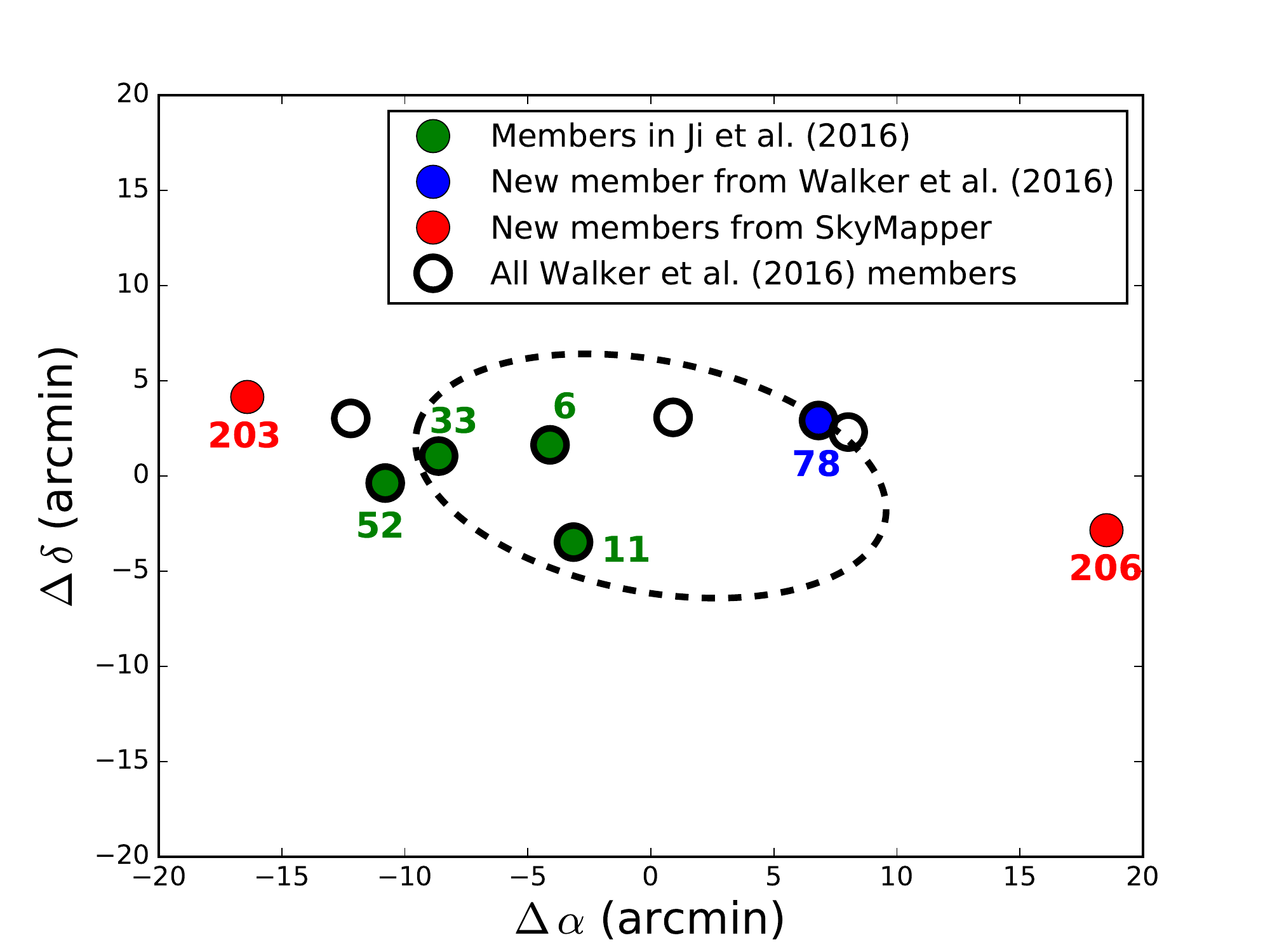}
\caption{Left: Color magnitude diagram of Tuc II stars from this study. 
A Dartmouth isochrone \citep{dcj+08} with an age of 12.5\,Gyr, distance modulus of 18.8, and a metallicity of $\mbox{[Fe/H]}= -2.5$ is overplotted in black along with offsets of (g$-$r)$\pm0.1$ in dashed lines to guide the eye.
We denote with different colors the four members previously observed by \citet{jfe+16}, one member selected from \citet{wmo+16}, and two selected from our SkyMapper photometry.
Open circles indicate confirmed members in \citet{wmo+16} with no high-resolution spectroscopic observations.
Right: Spatial distribution of Tuc II members centered on the coordinates of Tucana II. The elliptical half-light radius from \citet{kbt+15} is overplotted.
In both plots, each star is labeled by its identifier as found Table~\ref{tab:obs}.}
\label{fig:CMD}
\end{figure*}

The data in this paper were obtained with the Magellan Inamori Kyocera Echelle (MIKE) spectrograph on the Clay telescope at Las Campanas Observatory \citep{bsg+03}.
The observations were taken between August 14-17 and October 7-11, 2017.
Examples of the spectra are shown in Figure~\ref{fig:samplespec}. 
The location of each star in the color magnitude diagram of Tucana II is shown in Figure~\ref{fig:CMD}.
Targets were observed with 2x2 binning and the 1$\farcs$0 slit (R $\sim$ 28,000 on the blue chip and R $\sim$ 22,000 on the red chip) covering $\sim$3500\,\AA\,\,to $\sim9000$\,\AA.
The weather was mostly clear on all nights.
The spectra were all reduced and wavelength-calibrated with the MIKE CarPy pipeline\footnote{http://code.obs.carnegiescience.edu/mike} \citep{k+03}.

\section{Abundance Analysis}
\label{sec:abundances}

\begin{deluxetable}{lllll} 
\tablecolumns{5}
\tablewidth{\columnwidth}
\tablecaption{\label{tab:stellarparameters} Stellar parameters}
\tablehead{   
  \colhead{Name} &
  \colhead{$T_{\text{eff}}$} & 
  \colhead{Log\,$g$} &
  \colhead{v$_{\text{micro}}$} &
  \colhead{[Fe/H]} \\
  \colhead{} &
  \colhead{(K)} & 
  \colhead{(dex)} &
  \colhead{(km s$^{-1}$)} & 
  \colhead{(dex)}
}
\startdata
TucII-006 & $5017\pm227$ & $1.50\pm0.54$ & $1.95\pm0.24$ & $-2.93\pm0.27$ \\
TucII-011 & $4693\pm158$ & $1.25\pm0.50$ & $1.95\pm0.21$ & $-2.92\pm0.16$ \\
TucII-033 & $4828\pm153$ & $1.40\pm0.53$ & $1.80\pm0.21$ & $-2.41\pm0.12$ \\
TucII-052 & $4819\pm195$ & $1.70\pm0.42$ & $1.85\pm0.23$ & $-3.23\pm0.20$ \\
TucII-078 & $4954\pm178$ & $1.90\pm0.67$ & $2.20\pm0.23$ & $-2.69\pm0.20$ \\
TucII-203 & $4882\pm186$ & $1.60\pm0.33$ & $2.00\pm0.22$ & $-3.08\pm0.19$ \\
TucII-206 & $4900\pm186$ & $1.65\pm0.85$ & $1.90\pm0.26$ & $-3.34\pm0.28$\\
\enddata
\end{deluxetable}

\subsection{Derivation of stellar parameters and chemical abundances}
\label{sec:derivation}

The python-based Spectroscopy Made Hard (SMH) analysis software first described in  \citet{c+14} was used for the majority of our analysis, including normalizing spectra, measuring equivalent widths, and generating synthetic spectra.
Our version of this software made use of the 2011 version of MOOG \citep{s+73}, which has an updated treatment of scattering from \citet{sks+11}. The spectroscopic stellar parameter adjustment scheme by \citet{fcj+13} is based on this version. 
We used $\alpha$-enhanced, 1D plane-parallel stellar model atmospheres from \citet{ck+04}.
The line list in \citet{rpt+14} was used for identifying lines and deriving abundances from equivalent width measurements. 
For spectral syntheses, we supplemented this line list with those used in \citet{jfs+16}.
Namely, we incorporated lines from \citet{hpc+02}, \citet{dls+03}, \citet{iss+06}, \citet{lds+06}, \citet{lsc+09}, \citet{slc+09}, and \citet{mpv+14}.
Our chemical abundances are listed relative to the solar abundances of \citet{ags+09}.

We derived radial velocities by cross-correlating our observed spectra with a template spectrum of HD140283 over the H$\beta$ feature at 4861\,\AA.
Heliocentric velocity corrections were derived using the \texttt{rvcorrect} task in \texttt{IRAF}. 
We find evidence that TucII-078 may be in a binary, since our measured velocity is $\sim12$\,km/s greater than the velocity reported in \citet{wmo+16}.

We determined stellar parameters and chemical abundances following \citet{fcj+13} whose methodology we briefly outline in this paragraph.
First, equivalent widths were measured by fitting a Gaussian profile to each line.
We generally excluded lines with reduced equivalent width measurements greater than $-4.5$, since these measurements potentially lie outside the linear regime of the curve of growth.
We varied the stellar parameters ($T_{\text{eff}}$, $\log g$, $v_{\text{micro}}$, and [Fe/H]) until our Fe I abundances showed no trend with both excitation potential and reduced equivalent width.
We further constrained $\log g$ by requiring our Fe I and Fe II abundances to match.
We then corrected our $T_{\text{eff}}$ with the prescription given in \citet{fcj+13}, but re-adjusted $\log g$, $v_{\text{micro}}$, and [Fe/H] until the above criteria were again satisfied.
To determine random uncertainties, stellar parameters were varied to match the $1\sigma$ uncertainty in the Fe I abundance trends.
These random uncertainties were added in quadrature to the systematic uncertainties, which were assumed to be 150\,K for $T_{\text{eff}}$, 0.3\,dex for $\log g$, and 0.2\,km\,s$^{-1}$ for $v_{\text{micro}}$.
Final stellar parameter measurements are listed in Table~\ref{tab:stellarparameters}.

We followed a few prescriptions to determine uncertainties for abundances based on equivalent width measurements.
For abundances measured with a large number of lines (N $\geq$ 10), we take the standard deviation of the individual line abundances as the random uncertainty.
We adopt the standard deviation as it well represents abundance uncertainties obtained from data with poor signal to noise.
For abundances with a small number of lines (1 $<$ N $<$ 10), we derived random uncertainties by multiplying the range covered by the line abundances by the k-statistic following \citet{k+62} to obtain a standard deviation.
The k-statistic gives measurements with a smaller number of lines an appropriately larger uncertainty.
For abundances derived from only one line measurement, we derived the random uncertainty by varying the continuum placement and assuming the resulting abundance variation as the uncertainty.
If any resulting random uncertainty was below the standard deviation of the abundances of the iron lines, we nominally adopt as a conservative random uncertainty the standard deviation of the iron abundance (0.12\,dex to 0.27\,dex).
The total uncertainty for each element was then determined by adding the random uncertainty in quadrature with the systematic uncertainties.
The systematic uncertainties were assumed to be the difference in the abundances caused by varying each stellar parameter by its 1$\sigma$ uncertainty.

For abundances measured by spectrum syntheses, we also derived uncertainties by adopting the procedure in the previous paragraph.
If an element had only one synthesized line, the random uncertainty was assumed to be the change in abundance that was required to capture the variations of the continuum placement.
The systematic uncertainty was obtained by measuring the change in the abundance after varying each stellar parameter by its $1\sigma$ uncertainty.
If an element had measured abundances from both spectrum synthesis and equivalent width measurements, we pooled the measurements and derived random uncertainties following the procedure outlined in the previous paragraph.
The random uncertainty was then added in quadrature with the systematic uncertainties for each star to derive a total uncertainty.
Certain elements (e.g., Al and Si) had absorption features that were detected in our data, but the signal to noise was too poor to derive a meaningful abundance and especially uncertainty.
However, we report tentative abundances but mark them with a colon in Table~\ref{tab:abundancetable} to indicate a large uncertainty.
Our measurements and uncertainties are listed in Tables~\ref{tab:abundancetable} and~\ref{tab:uncertainties}.
Our individual equivalent width and synthesis measurements are listed in Table~\ref{tab:linetable}.

\subsection{Comparison to \citet{jfe+16} and \citet{wmo+16}}
\label{sec:comparison}

\begin{figure}[!htbp]
\centering
\includegraphics[width =\columnwidth]{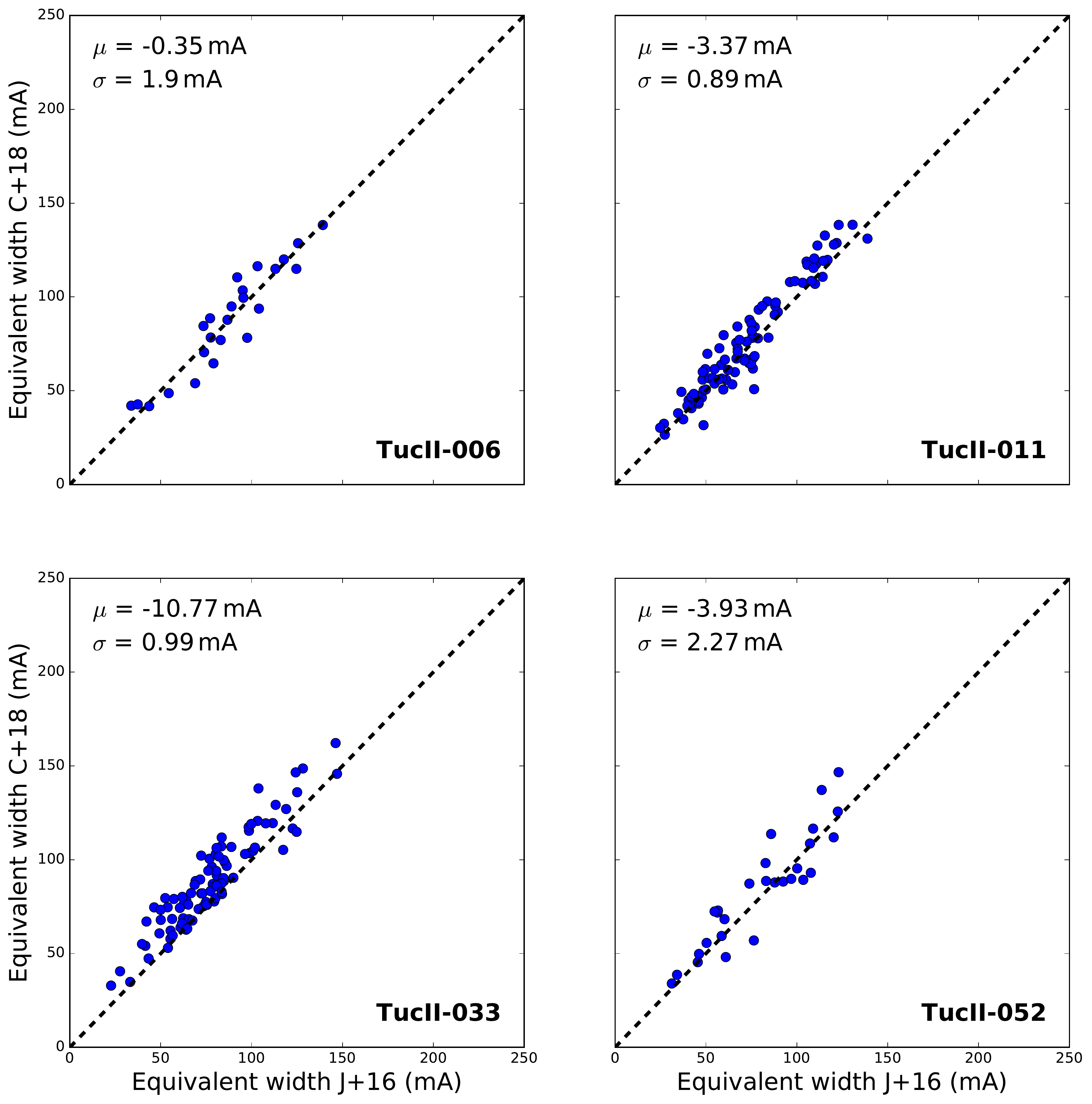}
\caption{Comparison of the equivalent widths of Fe I lines measured on the same spectra using our method and \citet{jfe+16}. The mean offset and standard error in the mean between our measurements are listed in each panel.}
\label{fig:eqwplot}
\end{figure}

We compare our results with measurements from \citet{jfe+16} and \citet{wmo+16} to check consistency with previous studies of Tucana\,II. 
We focus on comparisons with \citet{jfe+16}, with whom we have four stars in common, as they also analyzed high-resolution spectra from the MIKE spectrograph. 

We first discuss our measured stellar parameters and chemical abundances of TucII-006, TucII-011, TucII-033, and TucII-052 in comparison to those presented in \citet{jfe+16}.
For TucII-011 and TucII-052, we find excellent agreement (within 1$\sigma$) in stellar parameters and chemical abundances.
For TucII-006, we measure a discrepant $\log g$ by 0.4$\pm0.4$\,dex, a discrepant microturbulence by 0.25$\pm0.26$\,dex, and a discrepant [Fe/H] by 0.25$\pm0.21$\,dex, where the uncertainties are from \citet{jfe+16}.
However, 0.15\,dex of the discrepancy in [Fe/H] can be explained by differences in the stellar parameters, and the discrepancy in the stellar parameters can be explained by the lack of Fe II measurements for that star in \citet{jfe+16}. 
We measure three Fe II lines for the same star due to the better S/N of our spectra.
This comparison underscores the importance of propagating stellar parameter uncertainties to final abundance uncertainties, particularly in the case of spectra with low S/N and few lines.
For TucII-033, we measure a larger microturbulence and [Fe/H], which partially contributed to large discrepancies in the measurements of the Sr and Ba abundances (see Section~\ref{sec:tucii033}).
To isolate whether the discrepancies in measurements of TucII-033 were indeed due to the better S/N of our spectra, we performed our analysis on exactly the spectra used in \citet{jfe+16}.
Furthermore, we chose to analyze the spectra of all four stars in \citet{jfe+16} as a check on our method of measuring equivalent widths and deriving stellar parameters.

Applying our methodology to the same spectra that \citet{jfe+16} analyzed gives broadly consistent results.
We recover their measured $T_{\text{eff}}$ within their reported 1$\sigma$ bounds.
We also recover their $\log g$ measurements to within 1$\sigma$ for all stars.
We find general agreement within 2$\sigma$ between our microturbulence measurements and no obvious systematic effects.

All [Fe/H] measurements agree within 1$\sigma$ as well, but we measure a larger [Fe/H] by at least 0.15\,dex for three stars (TucII-006, TucII-033, and TucII-052).
For the star with the largest discrepancy (TucII-033), we thus inspected the equivalent width measurements. After inspecting fits to the individual absorption lines, it became apparent that the discrepancy is likely due to unfortunate continuum placement issues with the automated continuum fitting routine in previous work.
We thus inspected and compared equivalent widths of all stars, as shown in Figure~\ref{fig:eqwplot}. We find a small (somewhat) statistically significant difference between measurements for TucII-001 and TucII-052 but which are overall on the level of 3-4\,m{\AA}, and thus not a source for any significant abundance differences. No offset is found for TucII-006. For TucII-033, there is indeed a significant offset, of $\sim10$\,m{\AA}, which indeed explains why we measure systematically increased [Fe/H]. 

We conclude that any abundance discrepancies are consistent with previous uncertainties, but we now have significantly better S/N than before.
Thus, differences in our final stellar parameters and abundances between this work and \citet{jfe+16} are likely due to the additional observations that we have combined with theirs for a new analysis presented here.

\citet{wmo+16} measured metallicities for their Tucana II stars by matching their observed spectra of the Mg b region ($\sim$5150\,\AA) to a grid of synthetic spectra in the Segue Stellar Parameters Pipeline \citep[SSPP;][]{lss+08}.
They obtained $R\sim$18,000 spectra of their brighter targets and $R\sim$10,000 of their fainter targets.
We find that our metallicities are typically much lower (at least $\sim$0.50\,dex) than those in \citet{wmo+16} for the five stars in common, including the four stars in \citet{jfe+16}.
There is no obvious significant systematic difference between other stellar parameters that could explain this offset.
We do measure lower $T_{\text{eff}}$ values by 70\,K on average, but this difference does not explain such a large difference in the metallicities.
Neither do we see trends in our measured Mg abundances that could affect the Mg b region, and consequently, the metallicity measurements in \citet{wmo+16}.
Much of the discrepancy, however, can be attributed to the fact that \citet{wmo+16} applied a metallicity offset of $0.32$\,dex to their measurements based on an offset with respect to the measured metallicity of a solar spectrum, which may not be appropriate in our case, given that the Sun and these dwarf galaxy stars have very different stellar parameters.

\begin{figure*}[!htbp]
\centering
\includegraphics[width =0.99\textwidth]{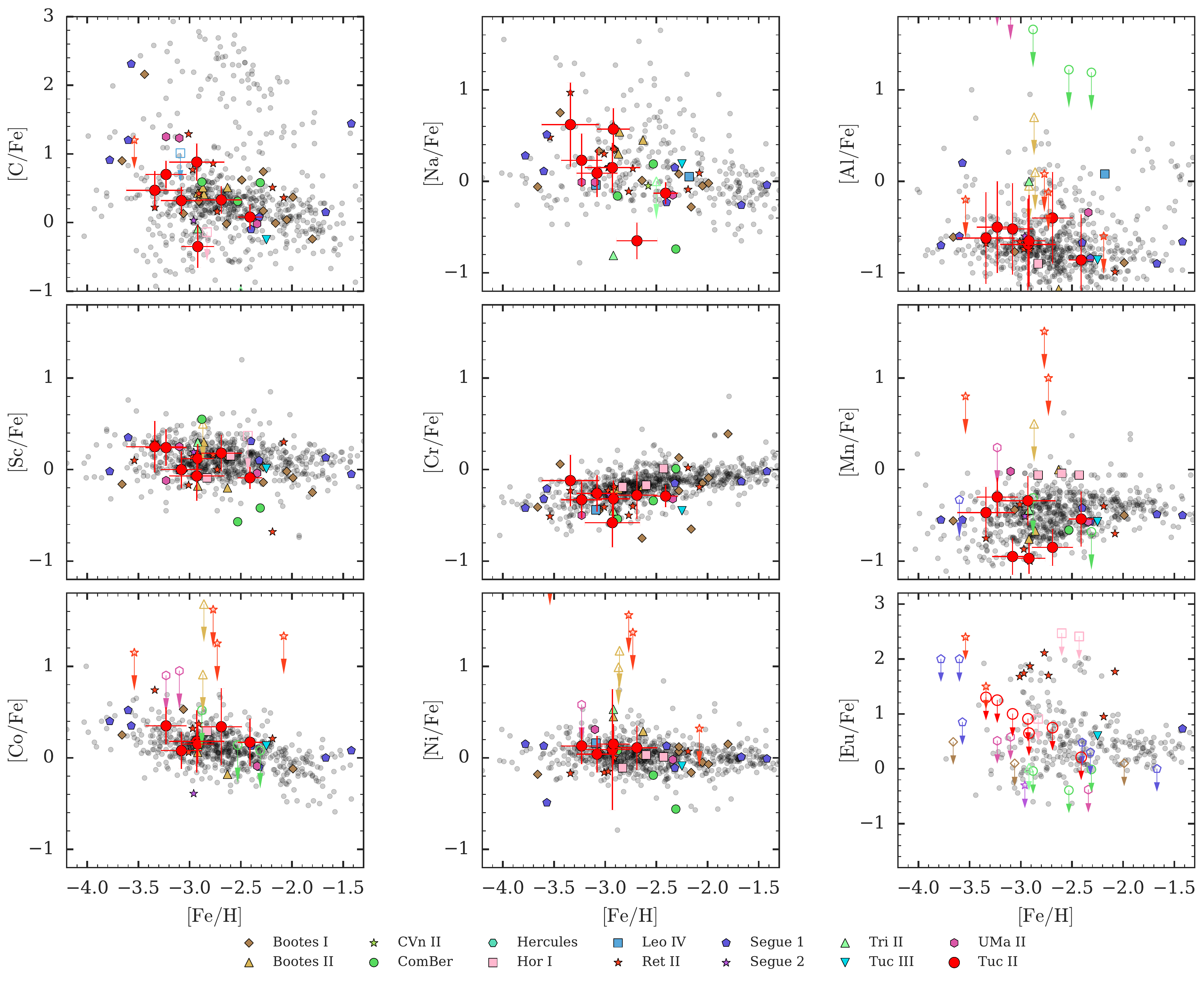}
\caption{[X/Fe] vs. [Fe/H] ratio for the abundances of carbon, the odd-Z elements, the iron-peak elements, and europium. 
Gray data points correspond to stars in the halo \citep{f+10, rpt+14}. 
Colored symbols are UFD stars. 
Error bars correspond to random uncertainties; see Table~\ref{tab:uncertainties} for total uncertainties.
Abundances marked by colons (:) in Table~\ref{tab:abundancetable} are shown with uncertainties of 0.5\,dex.
The carbon abundances in this plot are not corrected for the evolutionary state of each star following \citep{pfb+14}; see Table~\ref{tab:abundancetable} for corrected carbon abundances.
In general, the abundances of these elements in Tucana II stars agree with trends in other UFDs and the Milky Way halo.
UFD abundances are from \citet{kwg+08, fek+09, fks+10, fsg+10, fsk+14, ngw+10, nyg+10, jfw+10, gnm+13, kfa+13, iaa+14, kr+14, rk+14, jfe+16, jfs+16b, jfs+16, hsm+17, kcs+17, vsm+17, dms+17}.}
\label{fig:abplot1}
\end{figure*}

\begin{figure*}[!htbp]
\centering
\includegraphics[width =0.99\textwidth]{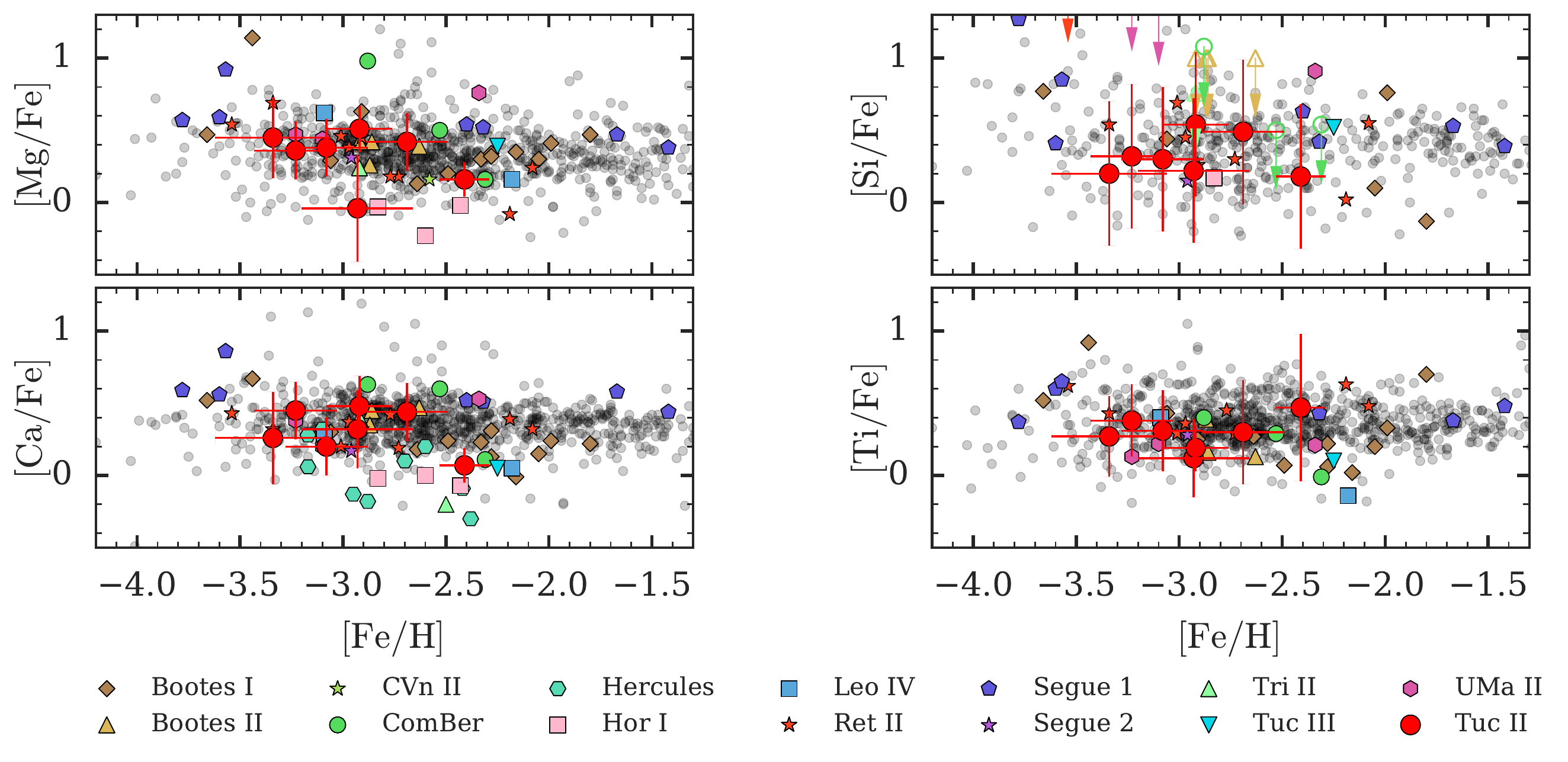}
\caption{[X/Fe] vs. [Fe/H] ratio of abundances of $\alpha$-element abundances in stars in Tucana II. 
Gray data points correspond to stars in the halo \citep{f+10, rpt+14}.  
Colored symbols are UFD stars. 
Error bars correspond to random uncertainties; see Table~\ref{tab:uncertainties} for total uncertainties.
Abundances marked by colons (:) in Table~\ref{tab:abundancetable} are shown with uncertainties of 0.5\,dex.
The decrease in the [$\alpha$/Fe] ratio of the most metal-rich star (TucII-033) would suggest that Tucana II had an extended star formation history, but see Section~\ref{sec:tucii033} for a discussion on the membership of TucII-033.}
\label{fig:abplot2}
\end{figure*}

\begin{figure*}[!htbp]
\centering
\includegraphics[width =0.99\textwidth]{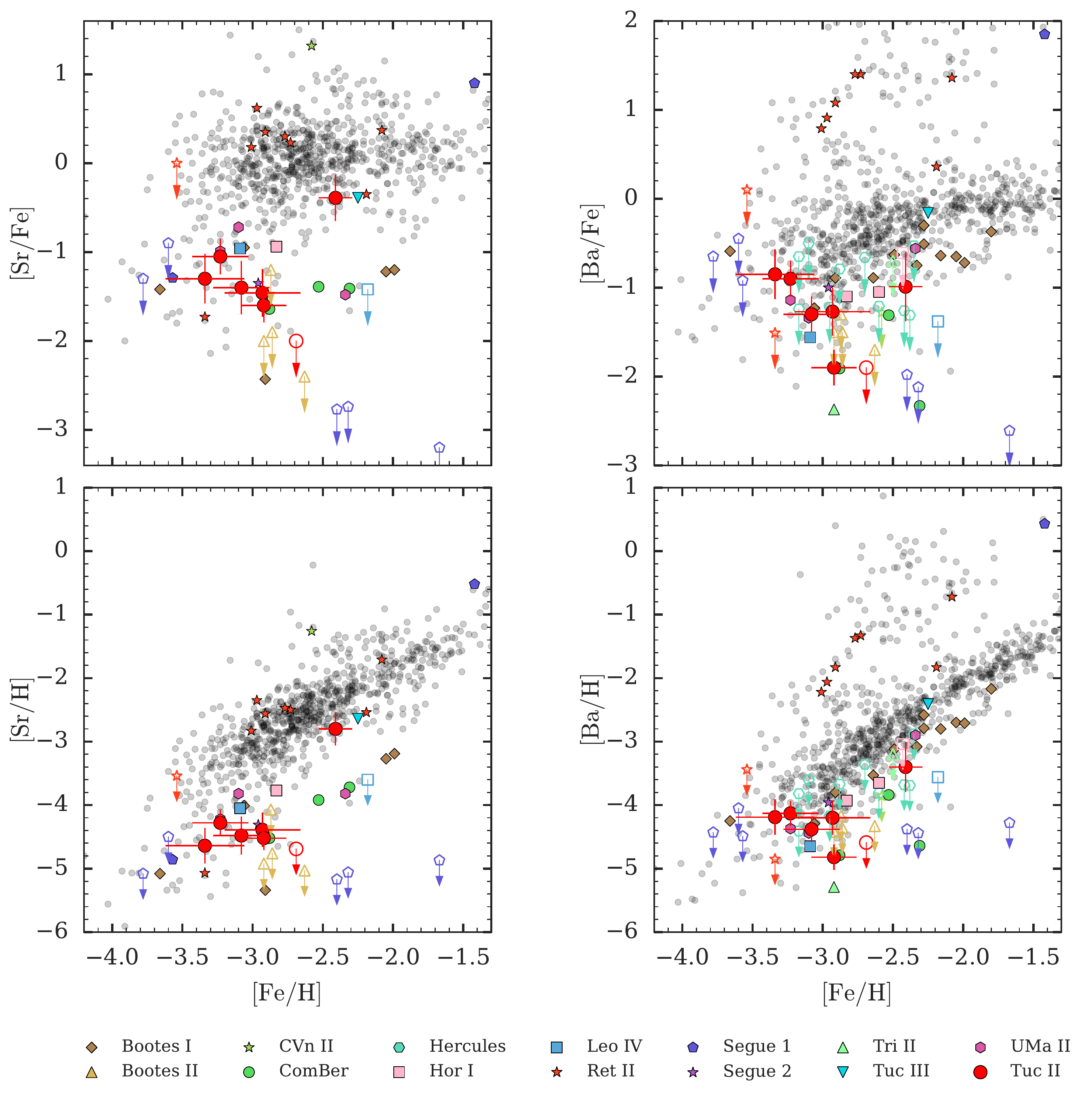}
\caption{[X/Fe] vs. [Fe/H] and [X/H] vs. [Fe/H] ratio of abundances of strontium and barium in stars in Tucana II. 
Gray data points correspond to stars in the halo \citep{f+10, rpt+14}. 
Colored symbols are UFD stars. 
Error bars correspond to random uncertainties; see Table~\ref{tab:uncertainties} for total uncertainties and Section~\ref{sec:derivation} for a discussion on deriving uncertainties.
The most metal-rich star in Tucana II (TucII-033) has Sr and Ba abundances that are above those typically seen in UFD stars.}
\label{fig:abplot3}
\end{figure*}

\section{Chemical signatures of the Tucana II stellar population}
\label{sec:signatures}

We first explore the possibility that TucII-033 is a halo interloper in Section~\ref{sec:tucii033}. Then, we discuss the trends of element abundances provided, and how they characterize the stellar population of Tucana II in the remainder of this section.

\subsection{Is TucII-033 a member of Tucana II?}
\label{sec:tucii033}

Traditionally, the membership status of stars in dwarf galaxies is derived from a combination of velocity and metallicity measurements.
However, the detailed chemical abundances of candidate member stars might also be used to determine membership because UFDs are expected to show distinct chemical signatures (e.g., lower Fe, Sr, Ba) compared to the halo background.
Additional evidence for non-membership might be gained if any star has chemical abundances distinct from those of other stars in the sample.
The small number of stars currently known in UFDs that do not necessarily yield well-defined abundance trends over large parameter space (e.g., [Fe/H]) requires, in particular, that any claim of chemical (non-)membership be investigated thoroughly.
In this section, we thus discuss if any stars in our sample have chemical signatures that challenge their radial velocity membership status.

All stars in our sample except one have abundances that are consistent with those of typical UFD stars, as can be seen in Figures~\ref{fig:abplot1},~\ref{fig:abplot2}, and~\ref{fig:abplot3}. 
The exception is TucII-033, the most metal-rich ([Fe/H] = $-2.41$) star.
It displays a Sr abundance ([Sr/Fe] = $-0.39$, [Sr/H] = $-2.8$) that is in disagreement with that of the typical UFD stars \citep{fsg+10,fsk+14}, and importantly, with that of the other stars in Tucana II. 
TucII-033 has a [Sr/H] abundance distinctly different by 1.7\,dex (a 50 fold increase) from five of the stars in Tucana\, II, which have an average [Sr/H] = $-4.46$ (with a standard deviation of only 0.14\,dex). 
The remaining star (TucII-078) is also distinct in that it has a low upper limit on its Sr abundance.
The lack of Sr in TucII-078 relative to the other Tucana\, II members is puzzling, but similar stars are known to exist in other UFDs (e.g., Segue I).
TucII-033 has an enhancement in Sr that appears to agree with the trend for halo stars as shown in Figure~\ref{fig:abplot3}, whereas TucII-078 and the other Tuc\,II members have Sr abundances far below the halo trend.
This comparison raises the possibility that TucII-033 might be an interloping halo star with the same systemic velocity as Tuc II.

To further investigate, we determined whether it was plausible for a halo star to have the same systemic velocity of Tucana II ($v_{\text{sys}} = -129.1$\,km/s; \citealt{wmo+16}).
We retrieved the velocities of halo stars with metallicities of [Fe/H] $< -2.0$ in the literature \citep{af+17}.
We find that this sample of 799 halo stars has a distribution of velocities that is roughly Gaussian and centered on 15\,km/s with a standard deviation of 154\,km/s. 
Using this distribution of velocities, we can calculate the odds of finding an interloping halo star around the mean systemic velocity of Tucana\,II. 
We derive a 6\% chance of finding a halo star within two times the velocity dispersion (8.6\,km/s in \citealt{wmo+16}) around the mean velocity of Tucana\,II, and a 9\% chance if we increase the bounds to three times the velocity dispersion. 
Thus, it is unlikely but not unreasonable for a metal-poor halo star to have the same systemic velocity as Tucana II. 
As an aside, we do note that considering exclusion from our sample likely does not affect the status of Tucana\, II as a dwarf galaxy.
\citet{wmo+16} measure a mean velocity for Tucana\,II of $-129^{+3.5}_{-3.5}$\,km/s and a velocity dispersion of $8.6^{+4.4}_{-2.7}$\,km/s. 
The velocity measurement of TucII-033 ($v_{\text{helio}} = -127.5$\,km/s) is close to 1$\sigma$ of the error on the measured systemic velocity of Tucana\, II. 
Thus, it is unlikely that the exclusion of this star would remove any velocity spread that is used to classify Tucana\, II as a UFD.

While there are also stars in Reticulum II \citep{jfs+16b}, a star in Tucana III \citep{hsm+17}, and a star in Canes Venatici II \citep{fmb+16} that show an enhancement in Sr, we do not consider them to be typical UFD stars.
In the case of Reticulum II and Tucana III, this Sr enhancement is reflective of strong and moderate $r$-process enrichments, respectively.
Given that the origin of these enhancements clearly derive from $r$-process events that occurred in these systems, and that these events are regarded rare, we do not consider them to be typical examples of UFDs.
Moreover, TucII-033 is not a $r$-process enhanced star.
It is difficult to judge the significance of the one available Sr abundance ([Sr/Fe] = 1.32) in Canes Venatici II.
This star has a Sr enhancement that could be a result of a weak r-process enrichment event \citep[e.g.,][]{w+13} and in theory, a similar event may have enhanced TucII-033.
However, more data from Canes Venatici II is needed to derive firm conclusions.
Thus, around the metallicity of TucII-033 ([Fe/H] $\sim -$2.5), the typical UFD stellar population either has extremely low upper limits on the Sr abundance (i.e., Segue 1; [Sr/H] $\lesssim -4.0$) or marginal detections (e.g., stars in Coma Berenices, Ursa Major II; \citealt{fsg+10, fsk+14}).

This high Sr abundance measurement of TucII-033 also naturally raises the question of why the Sr abundance was not recognized as such in the previous study of this galaxy.
Upon investigation, we find that we measure a higher Sr abundance than \citet{jfe+16} by 0.70\,dex.
However, their Sr abundance has a large uncertainty ($\sim0.6$\,dex) and somewhat distorted lines due to low S/N at the Sr lines (4077\AA\, and 4215\AA).
Our improved S/N in this region clearly shows that high Sr is required. 

We do note that TucII-033 is distinct from halo stars in that it has a markedly lower [$\alpha$/Fe] ratio ($\sim$0.05\,dex) than other halo stars ($\sim0.4$\,dex) as discussed in Section~\ref{sec:alpha}.
Using the compilation by \citet{af+17}, we find that only 8\% of halo stars have a lower Ca abundance than TucII-033 and 15\% have a lower Mg abundance.
These fractions, when viewed in the context that TucII-033 has the same systemic velocity as Tucana II, make it less likely that TucII-033 is an interloping star.

For these reasons, for the remainder of the analysis, we present two lines of argument: one assuming TucII-033 as a member, and one assuming TucII-033 as a non-member. 
The exclusion of TucII-033 from the interpretation of this galaxy would be meaningful, since its low [$\alpha$/Fe] abundance would otherwise imply that Tucana II had an extended star formation history and would thus not be a surviving first galaxy (see Sections~\ref{sec:alpha} and~\ref{sec:firstgalaxy}).
While the Sr abundance of TucII-033 might suggest that it is a halo interloper, its [$\alpha$/Fe] and velocity make this scenario less likely.

\subsection{Carbon}
\label{sec:carbon}
Empirically, a high fraction of EMP stars in the halo ($\sim42\%$; \citet{pfb+14}) are enhanced in carbon ([C/Fe] $>$ 0.7\,dex) and are thus classified as carbon-enhanced metal-poor stars.
This enhancement in carbon has been used to constrain potential sites of nucleosynthesis that may have dominated early chemical evolution \citep[e.g.,][]{tmu+07, cm+14, fn+15}.
From the paradigm of hierarchical galaxy formation, we might expect that stars in dwarf galaxies also display this enhancement given that accreted analogs perhaps contributed to the metal-poor population of the halo. 
However, recent studies of the prevalence of carbon-rich stars in dwarf galaxies give differing results \citep{kgz+15, jnm+15, csf+18}.

In Tucana II, we find that three stars out of five with $\mbox{[Fe/H]}<-2.9$ are enhanced in carbon, following the correction in \citet{pfb+14}.
This fraction is slightly larger than that of the halo, with the caveat of the small size of our sample.
One star (TucII-011) appears to be somewhat less enhanced in carbon ([C/Fe]$=0.29$ after correction for the evolutionary state of the star). 
This slight outlier might reflect inhomogeneous mixing of gas in the system or multiple avenues of enrichment that contributed to the chemical evolution of the system.

\subsection{$\alpha$-elements}
\label{sec:alpha}

The abundance of $\alpha$-elements (Mg, Si, Ca, Ti) in stars can be used investigate the integrated population of supernovae (SNe) that chemically enriched the natal gas cloud of the stars.
In particular, enrichment by core-collapse  SNe results in a flat [$\alpha$/Fe]$\,\,\sim 0.4$ trend vs. [Fe/H], whereas Type Ia SNe enrichment result in a declining [$\alpha$/Fe] abundance trend vs. [Fe/H] \citep[e.g.][]{ibn+99, kcs+11}. 
Typically, this switch from a flat to a declining [$\alpha$/Fe] vs. [Fe/H] indicates the metallicity at which type Ia SNe started dominating the Fe production. 
The most metal-rich member of our sample, TucII-033, shows a slight deficiency in its $\alpha$-element abundance compared to the other stars in our sample. 
This deficiency suggests that Type Ia SNe contributed to the chemical abundances of at least some stars in this galaxy, which in turn would suggest somewhat extended star formation and chemical enrichment in Tucana II \citep[e.g.,][]{kcs+11}.
Declining  [$\alpha$/Fe] is seen in most UFDs \citep{vgk+13}, though Tucana II is one of the least luminous UFDs with available [$\alpha$/Fe] measurements.

If TucII-033 were a halo interloper (see Section~\ref{sec:tucii033}), our sample would instead show a trend consistent with a constant [$\alpha$/Fe], as produced by core-collapse SNe only. 
We note, that the lower Mg abundance displayed by TucII-006 is likely due to a distortion in one of the two lines used to measure its abundance, which is reflected in the larger uncertainties on its Mg abundance of 0.46\,dex.

\subsection{Odd-Z, iron-peak and neutron-capture element abundances}
\label{sec:otherchem}

We find no significant deviation from the halo trend in the odd-Z elements (Na, Al, Sc) and iron-peak elements (Cr, Mn, Co, Ni).
This is also consistent with abundances of other UFD stars in the literature (see Figures~\ref{fig:abplot1} and~\ref{fig:abplot2}).
We do find one star (TucII-078) with a lower Na abundance than typical halo and dwarf galaxy stars. 

We find low strontium and barium abundances in six stars that are characteristic of the stellar populations set by other UFD stars (see Figure~\ref{fig:abplot3}).
See Section~\ref{sec:tucii033} for a discussion of the neutron-capture element abundances in TucII-033.
We do not detect europium or other neutron-capture elements in any stars in our sample.

\subsection{Tucana II as a surviving first galaxy}
\label{sec:firstgalaxy}

\citet{fb+12} predict chemical characteristics of the population of surviving first galaxies: 

\begin{enumerate}
\item A large spread in [Fe/H] ($\sim1\,$dex), 
\item Light element abundance ratios in agreement with a core-collapse supernova enrichment
\item No stars with [$\alpha$/Fe] systemically lower than the galactic halo abundance of [$\alpha$/Fe] $\sim$ 0.35, and 
\item No signatures of $s$-process enhancement from AGB stars.
\end{enumerate}

The first prediction is a consequence of inhomogeneous mixing in the first galaxies, and the last two predictions are consequences of the first galaxies being enriched by ``one-shot'' chemical enrichment events since extended star formation is not thought to have occurred in the first galaxies. The second criterion simply confirms enrichment by supernovae. According to our chemical abundance results, Tucana II largely satisfies the aforementioned first, second and fourth criteria for a surviving first galaxy.
The second one would also be satisfied if we exclude TucII-033 when assuming it is a halo interloper (see Section~\ref{sec:tucii033}).
However, if we include TucII-033 in the interpretation, then its Sr relative enhancement relative to other Tuc\,II members and [$\alpha$/Fe] deficiency would imply that Tuc\,II had undergone some period of chemical evolution, and would thus not be a surviving first galaxy. 
Given the low luminosity of Tucana II ($\sim3 \times 10^3\,\text{L}_\odot$; \citealt{bdb+15, kbt+15}), it would still be interesting to find that Tucana II has some chemical evolution as opposed to isolated chemical enrichment events.

Together with Segue\,1 \citep{fsk+14}, Tucana II might still be one of the best candidates for a surviving first galaxy, as determined from chemical abundances of six stars in each galaxy.
Moreover, these six stars form the majority of known members in Tucana II, and their chemical abundances suggest that most stars in the galaxy are consistent with having formed in a first galaxy environment. 
Theoretical modeling of early galaxies such as these two systems could shed further light on this issue.
However, detailed abundances of more stars with [Fe/H]$\gtrsim -2.5$ in Tucana II are needed to further investigate the nature and origin of Tucana II, as is the case with other potential first galaxy candidates (i.e., Ursa Major II, Coma Berenices, Leo IV).

\section{Conclusion}
\label{sec:conclusions}

In this paper, we presented the high-resolution chemical abundance measurements of seven stars in the Tucana II dwarf galaxy.
Three stars with no previous high-resolution chemical abundance measurements were analyzed.
Four other stars had been reanalyzed from the sample in \citet{jfe+16} with the addition of new data.

From the detection of new members and the re-analysis of known members, we were able to discuss the chemical signatures of stars in the Tucana II UFD. 
We raise the possibility that one of the stars (TucII-033) may be an interloping halo star given its high Sr abundance with respect to other known UFD stars, but its velocity and [$\alpha$/Fe] ratio make this unlikely.
Excluding TucII-033 from the interpretation, Tucana II does meet all the criteria to be a surviving first galaxy \citep{fb+12}.
Assuming TucII-033 is a member, Tucana II would not meet the one-shot enrichment criterion due to the star's low [$\alpha$/Fe] and likely somewhat extended chemical evolution.

We confirmed two new members of Tucana II that were pre-selected as probable members from SkyMapper photometry.
Given the large field of view of the SkyMapper telescope (5.7 sq. deg.) and the metallicity discriminating ``$v$" filter, we were able to search for metal-poor stars within a large area around Tuc II UFD.
As a result, our two new members are $\sim2$ half-light radii from the center of Tucana II and may have been missed by traditional spectroscopic follow-up observations (see Figure~\ref{fig:CMD}).
Interestingly, one of these new members is the most metal-poor star discovered in Tucana II thus far ([Fe/H] = $-3.34$).
From our small sample, we however cannot claim these new members display systematic differences to stars near the center of Tucana II.

This new photometric metal-poor star identification technique might aid in identifying members for detailed chemical analysis and studying potential correlations with substructure of UFD systems.
Combining this photometric selection technique with traditional spectroscopic follow-up would result in more accurate parameters for UFDs (e.g., half-light radii, mass-to-light ratios), supposing the photometry itself could predict membership status.
In particular, this highly efficient large-field of view method for finding members would be interesting to apply on systems that show potential elongated tidal features, such as Tucana III \citep{dba+16,sld+17}.
Moreover, the spectroscopic study of more stars in UFDs would  have multiple benefits. For instance, detecting more stars with [Fe/H]$\sim-2.5$ in Tuc\,II would potentially resolve whether the entire galaxy had undergone chemical evolution. This would better inform whether TucII-033 is indeed a halo interloper or rather signaling unusual enrichment events or some degree of chemical evolution in that UFD.
At minimum, future work will extend this selection technique to other UFDs for the purpose of efficiently identifying new members and enabling detailed abundance measurements.

\acknowledgements

We thank Dougal Mackey and Christian Wolf for helpful comments on reducing the SkyMapper photometry, and we thank Vinicius Placco for computing carbon corrections.
A.C. and A.F. are supported by NSF CAREER grant AST-1255160.
A.F. acknowledges partial support from PHY 14-30152; and Physics Frontier Center/JINA Center for the
Evolution of the Elements (JINA-CEE), awarded by the
US National Science Foundation. 
APJ is supported by NASA through Hubble Fellowship grant HST-HF2-51393.001 awarded by the Space Telescope Science Institute, which is operated by the Association of Universities for Research in Astronomy, Inc., for NASA, under contract NAS5-26555.
HJ and JEN acknowledge the support of the
Australian Research Council through Discovery project DP150100862.
The national facility capability for SkyMapper has been funded through ARC LIEF grant LE130100104 from the Australian Research Council, awarded to the University of Sydney, the Australian National University, Swinburne University of Technology, the University of Queensland, the University of Western Australia, the University of Melbourne, Curtin University of Technology, Monash University and the Australian Astronomical Observatory. SkyMapper is owned and operated by The Australian National University's Research School of Astronomy and Astrophysics.
This work made use
of NASA’s Astrophysics Data System Bibliographic Services. 
This work has also made extensive use of the astropy package \citep{astropy}.

Facilities: Magellan-Clay (MIKE), SkyMapper

\software{Turbospectrum \citep{ap+98, p+12}, MARCS \citep{gee+08}, MIKE CarPy \citep{k+03}, MOOG \citep{s+73}, Astropy \citep{astropy}}

\bibliography{tucii}

\newpage

\startlongtable
\begin{deluxetable*}{lrrrrr|@{\hskip 0.1in}lrrrrrr} 
\tablecaption{$\sigma$ values correspond to random uncertainties. 
See Table~\ref{tab:uncertainties} for total uncertainties.
Colons indicate measurements with large uncertainties.}
\tablecolumns{12}
\tablewidth{\textwidth}
\tablecaption{Chemical abundances\label{tab:abundancetable}}
\tablehead{   
  \colhead{El.} &
  \colhead{N} &
  \colhead{$\log\epsilon(\text{X})_{\odot}$} &
  \colhead{[X/H]} & 
  \colhead{[X/Fe]} & 
  \colhead{$\sigma\tablenotemark{a}$} {\hskip 0.05in} &
  \colhead{El.} &
  \colhead{N} &
  \colhead{$\log\epsilon(\text{X})_{\odot}$} &
  \colhead{[X/H]} & 
  \colhead{[X/Fe]} & 
  \colhead{$\sigma\tablenotemark{a}$} 
}
\startdata
\multicolumn{6}{c}{\textbf{TucII-006}} &
\multicolumn{6}{c}{\textbf{TucII-011}}\\
\bottomrule\\
CH  &  2  & 8.43 & $-$2.05 & 0.88 & 0.27 {\hskip 0.05in} & {\hskip 0.05in} CH  &  2  & 8.43 & $-$3.27 & $-$0.35 & 0.31 \\ 
CH\tablenotemark{b}  &  2  & 8.43 & $-$1.69 & 1.24 & 0.27 {\hskip 0.05in} & {\hskip 0.05in} CH\tablenotemark{b}  &  2  & 8.43 & $-$2.63 & 0.29 & 0.31  \\ 
Na I  &  2  & 6.24 & $-$2.78 & 0.15 & 0.27 {\hskip 0.05in} & {\hskip 0.05in} Na I  &  2  & 6.24 & $-$2.35 & 0.57 & 0.23 \\ 
Mg I  &  2  & 7.60 & $-$2.97 & $-$0.04 & 0.37 {\hskip 0.05in} & {\hskip 0.05in} Mg I  &  5  & 7.60 & $-$2.41 & 0.51 & 0.16 \\ 
Al I  &  1  & 6.45 & $-$3.62:\tablenotemark{c} & $-$0.69:\tablenotemark{c} & \nodata {\hskip 0.05in} & {\hskip 0.05in} Al I  &  2  & 6.45 & $-$3.57:\tablenotemark{c} & $-$0.65:\tablenotemark{c} & \nodata \\ 
Si I  &  1  & 7.51 & $-$2.71:\tablenotemark{c} & 0.22:\tablenotemark{c} & \nodata {\hskip 0.05in} & {\hskip 0.05in} Si I  &  2  & 7.51 & $-$2.38:\tablenotemark{c} & 0.54:\tablenotemark{c} & \nodata \\ 
Ca I  &  2  & 6.34 & $-$2.61 & 0.32 & 0.27 {\hskip 0.05in} & {\hskip 0.05in} Ca I  &  10  & 6.34 & $-$2.44 & 0.48 & 0.21 \\ 
Sc II  &  5  & 3.15 & $-$3.00 & $-$0.07 & 0.27 {\hskip 0.05in} & {\hskip 0.05in} Sc II  &  6  & 3.15 & $-$2.80 & 0.12 & 0.16 \\ 
Ti II  &  12  & 4.95 & $-$2.81 & 0.12 & 0.27 {\hskip 0.05in} & {\hskip 0.05in} Ti I  &  2  & 4.95 & $-$2.73 & 0.19 & 0.16 \\ 
Cr I  &  3  & 5.64 & $-$3.51 & $-$0.58 & 0.27 {\hskip 0.05in} & {\hskip 0.05in} Ti II  &  17  & 4.95 & $-$2.60 & 0.32 & 0.22 \\ 
Mn I  &  3  & 5.43 & $-$3.27 & $-$0.34 & 0.27 {\hskip 0.05in} & {\hskip 0.05in} Cr I  &  5  & 5.64 & $-$3.24 & $-$0.32 & 0.20 \\ 
Fe I  &  52  & 7.50 & $-$2.93 & 0.00 & 0.27 {\hskip 0.05in} & {\hskip 0.05in} Mn I  &  3  & 5.43 & $-$3.89 & $-$0.97 & 0.17 \\ 
Fe II  &  3  & 7.50 & $-$2.93 & 0.00 & 0.27 {\hskip 0.05in} & {\hskip 0.05in} Fe I  &  99  & 7.50 & $-$2.92 & 0.00 & 0.16 \\ 
Co I  &  1  & 4.99 & $-$2.75 & 0.18 & 0.34 {\hskip 0.05in} & {\hskip 0.05in} Fe II  &  11  & 7.50 & $-$2.91 & 0.01 & 0.16 \\ 
Ni I  &  2  & 6.22 & $-$2.84 & 0.09 & 0.66 {\hskip 0.05in} & {\hskip 0.05in} Co I  &  3  & 4.99 & $-$2.77 & 0.15 & 0.21 \\ 
Sr II  &  2  & 2.87 & $-$4.39 & $-$1.46 & 0.27 {\hskip 0.05in} & {\hskip 0.05in} Ni I  &  3  & 6.22 & $-$2.77 & 0.15 & 0.22 \\ 
Ba II  &  2  & 2.18 & $-$4.20 & $-$1.27 & 0.27 {\hskip 0.05in} & {\hskip 0.05in} Sr II  &  2  & 2.87 & $-$4.52 & $-$1.60 & 0.19 \\ 
Eu I  &  1  & 0.52 & $<-$2.02 & $<$0.91 & \nodata {\hskip 0.05in} & {\hskip 0.05in} Ba II  &  1  & 2.18 & $-$4.82 & $-$1.90 & 0.20 \\ 
& & & & & {\hskip 0.05in} & {\hskip 0.05in} Eu I  &  1  & 0.52 & $<-$2.27 & $<$0.65 & \nodata \\ 
\bottomrule\\
\multicolumn{6}{c}{\textbf{TucII-033}} &
\multicolumn{6}{c}{\textbf{TucII-052}}\\
\bottomrule\\
CH  &  2  & 8.43 & $-$2.33 & 0.08 & 0.18 {\hskip 0.05in} & {\hskip 0.05in} CH  &  2  & 8.43 & $-$2.53 & 0.70 & 0.20 \\ 
CH\tablenotemark{b}  &  2  & 8.43 & $-$1.81 & 0.60 & 0.18 {\hskip 0.05in} & {\hskip 0.05in} CH\tablenotemark{b}  &  2  & 8.43 & $-$2.33 & 0.90 & 0.20 \\ 
Na I  &  2  & 6.24 & $-$2.54 & $-$0.13 & 0.12 {\hskip 0.05in} & {\hskip 0.05in} Na I  &  2  & 6.24 & $-$3.00 & 0.23 & 0.29 \\ 
Mg I  &  4  & 7.60 & $-$2.25 & 0.16 & 0.12 {\hskip 0.05in} & {\hskip 0.05in} Mg I  &  2  & 7.60 & $-$2.87 & 0.36 & 0.20 \\ 
Al I  &  2  & 6.45 & $-$3.27:\tablenotemark{c} & $-$0.86:\tablenotemark{c} & \nodata {\hskip 0.05in} & {\hskip 0.05in} Al I  &  1  & 6.45 & $-$3.73:\tablenotemark{c} & $-$0.50:\tablenotemark{c} & \nodata \\ 
Si I  &  2  & 7.51 & $-$2.23:\tablenotemark{c} & 0.18:\tablenotemark{c} & \nodata {\hskip 0.05in} & {\hskip 0.05in} Si I  &  2  & 7.51 & $-$2.91:\tablenotemark{c} & 0.32:\tablenotemark{c} & \nodata \\ 
Ca I  &  8  & 6.34 & $-$2.34 & 0.07 & 0.12 {\hskip 0.05in} & {\hskip 0.05in} Ca I  &  4  & 6.34 & $-$2.78 & 0.45 & 0.20 \\ 
Sc II  &  5  & 3.15 & $-$2.50 & $-$0.09 & 0.12 {\hskip 0.05in} & {\hskip 0.05in} Sc II  &  5  & 3.15 & $-$2.99 & 0.24 & 0.20 \\ 
Ti I  &  2  & 4.95 & $-$1.94 & 0.47 & 0.51 {\hskip 0.05in} & {\hskip 0.05in} Ti II  &  12  & 4.95 & $-$2.85 & 0.38 & 0.25 \\ 
Ti II  &  12  & 4.95 & $-$2.49 & $-$0.08 & 0.27 {\hskip 0.05in} & {\hskip 0.05in} Cr I  &  5  & 5.64 & $-$3.56 & $-$0.33 & 0.20 \\ 
Cr II  &  1  & 5.64 & $-$2.42 & $-$0.01 & 0.32 {\hskip 0.05in} & {\hskip 0.05in} Mn I  &  4  & 5.43 & $-$3.53 & $-$0.30 & 0.21 \\ 
Cr I  &  5  & 5.64 & $-$2.70 & $-$0.29 & 0.12 {\hskip 0.05in} & {\hskip 0.05in} Fe I  &  56  & 7.50 & $-$3.23 & 0.00 & 0.20 \\ 
Mn I  &  3  & 5.43 & $-$2.95 & $-$0.54 & 0.30 {\hskip 0.05in} & {\hskip 0.05in} Fe II  &  3  & 7.50 & $-$3.21 & 0.02 & 0.20 \\ 
Fe I  &  92  & 7.50 & $-$2.41 & 0.00 & 0.12 {\hskip 0.05in} & {\hskip 0.05in} Co I  &  3  & 4.99 & $-$2.88 & 0.35 & 0.20 \\ 
Fe II  &  12  & 7.50 & $-$2.39 & 0.02 & 0.14 {\hskip 0.05in} & {\hskip 0.05in} Ni I  &  2  & 6.22 & $-$3.10 & 0.13 & 0.20 \\ 
Co I  &  2  & 4.99 & $-$2.24 & 0.17 & 0.26 {\hskip 0.05in} & {\hskip 0.05in} Sr II  &  2  & 2.87 & $-$4.28 & $-$1.05 & 0.20 \\ 
Sr II  &  2  & 2.87 & $-$2.80 & $-$0.39 & 0.26 {\hskip 0.05in} & {\hskip 0.05in} Ba II  &  2  & 2.18 & $-$4.13 & $-$0.90 & 0.20 \\ 
Ba II  &  2  & 2.18 & $-$3.40 & $-$0.99 & 0.39 {\hskip 0.05in} & {\hskip 0.05in} Eu I  &  1  & 0.52 & $<-$1.98 & $<$1.25 & \nodata \\ 
Eu I  &  1  & 0.52 & $<-$2.21 & $<$0.20 & \nodata {\hskip 0.05in} & {\hskip 0.05in} & & & & & &\\ 
\bottomrule\\
\multicolumn{6}{c}{\textbf{TucII-078}} &
\multicolumn{6}{c}{\textbf{TucII-203}}\\
\bottomrule\\
CH  &  2  & 8.43 & $-$2.36 & 0.33 & 0.20 {\hskip 0.05in} & {\hskip 0.05in} CH  &  2  & 8.43 & $-$2.76 & 0.32 & 0.20 \\ 
CH\tablenotemark{b}  &  2  & 8.43 & $-$2.26 & 0.43 & 0.20 {\hskip 0.05in} & {\hskip 0.05in} CH\tablenotemark{b}  &  2  & 8.43 & $-$2.44 & 0.64 & 0.20 \\ 
Na I  &  2  & 6.24 & $-$3.34 & $-$0.65 & 0.20 {\hskip 0.05in} & {\hskip 0.05in} Na I  &  2  & 6.24 & $-$2.99 & 0.09 & 0.26 \\ 
Mg I  &  5  & 7.60 & $-$2.27 & 0.42 & 0.20 {\hskip 0.05in} & {\hskip 0.05in} Mg I  &  2  & 7.60 & $-$2.70 & 0.38 & 0.20 \\ 
Al I  &  2  & 6.45 & $-$3.09:\tablenotemark{c} & $-$0.40:\tablenotemark{c} & \nodata {\hskip 0.05in} & {\hskip 0.05in} Al I  &  2  & 6.45 & $-$3.60:\tablenotemark{c} & $-$0.52:\tablenotemark{c} & \nodata \\ 
Si I  &  2  & 7.51 & $-$2.20:\tablenotemark{c} & 0.49:\tablenotemark{c} & \nodata {\hskip 0.05in} & {\hskip 0.05in} Si I  &  1  & 7.51 & $-$2.78:\tablenotemark{c} & 0.30:\tablenotemark{c} & \nodata \\ 
Ca I  &  8  & 6.34 & $-$2.25 & 0.44 & 0.20 {\hskip 0.05in} & {\hskip 0.05in} Ca I  &  4  & 6.34 & $-$2.88 & 0.20 & 0.20 \\ 
Sc II  &  6  & 3.15 & $-$2.51 & 0.18 & 0.20 {\hskip 0.05in} & {\hskip 0.05in} Sc II  &  5  & 3.15 & $-$3.08 & 0.00 & 0.20 \\ 
Ti I  &  1  & 4.95 & $-$2.39 & 0.30 & 0.36 {\hskip 0.05in} & {\hskip 0.05in} Ti II  &  14  & 4.95 & $-$2.77 & 0.31 & 0.28 \\ 
Ti II  &  20  & 4.95 & $-$2.22 & 0.47 & 0.25 {\hskip 0.05in} & {\hskip 0.05in} V II  &  1  & 3.93 & $-$1.88 & 1.20 & 0.21 \\ 
Cr I  &  5  & 5.64 & $-$2.97 & $-$0.28 & 0.26 {\hskip 0.05in} & {\hskip 0.05in} Cr I  &  3  & 5.64 & $-$3.34 & $-$0.26 & 0.20 \\ 
Mn I  &  2  & 5.43 & $-$3.54 & $-$0.85 & 0.20 {\hskip 0.05in} & {\hskip 0.05in} Mn I  &  3  & 5.43 & $-$4.03 & $-$0.95 & 0.20 \\ 
Fe I  &  83  & 7.50 & $-$2.69 & 0.00 & 0.20 {\hskip 0.05in} & {\hskip 0.05in} Fe I  &  59  & 7.50 & $-$3.08 & 0.00 & 0.20 \\ 
Fe II  &  9  & 7.50 & $-$2.69 & 0.00 & 0.20 {\hskip 0.05in} & {\hskip 0.05in} Fe II  &  4  & 7.50 & $-$3.07 & 0.01 & 0.20 \\ 
Co I  &  1  & 4.99 & $-$2.35 & 0.34 & 0.42 {\hskip 0.05in} & {\hskip 0.05in} Co I  &  2  & 4.99 & $-$3.00 & 0.08 & 0.20 \\ 
Ni I  &  2  & 6.22 & $-$2.58 & 0.11 & 0.24 {\hskip 0.05in} & {\hskip 0.05in} Ni I  &  3  & 6.22 & $-$3.04 & 0.04 & 0.20 \\ 
Sr II  &  1  & 2.87 & $<-$4.69 & $<-$2.00 & \nodata{\hskip 0.05in} & {\hskip 0.05in} Sr II  &  1  & 2.87 & $-$4.48 & $-$1.40 & 0.30 \\ 
Ba II  &  1  & 2.18 & $<-$4.59 & $<-$1.90 & \nodata {\hskip 0.05in} & {\hskip 0.05in} Ba II  &  1  & 2.18 & $-$4.38 & $-$1.30 & 0.20 \\ 
Eu I  &  1  & 0.52 & $<-$1.94 & $<$0.75 & \nodata {\hskip 0.05in} & {\hskip 0.05in} Eu I  &  1  & 0.52 & $<-$2.08 & $<$1.00 & \nodata \\ 
\bottomrule\\
\multicolumn{6}{c}{\textbf{TucII-206}}\\
\bottomrule\\
CH  &  2  & 8.43 & $-$2.87 & 0.47 & 0.28 {\hskip 0.05in}\\ 
CH\tablenotemark{b}  &  2  & 8.43 & $-$2.61 & 0.73 & 0.28 {\hskip 0.05in}\\
Na I  &  2  & 6.24 & $-$2.72 & 0.62 & 0.46 {\hskip 0.05in}\\ 
Mg I  &  3  & 7.60 & $-$2.89 & 0.45 & 0.28 {\hskip 0.05in}\\ 
Al I  &  2  & 6.45 & $-$3.96:\tablenotemark{c} & $-$0.62:\tablenotemark{c} & \nodata {\hskip 0.05in}\\ 
Si I  &  1  & 7.51 & $-$3.14:\tablenotemark{c} & 0.20:\tablenotemark{c} & \nodata {\hskip 0.05in}\\ 
Ca I  &  2  & 6.34 & $-$3.08 & 0.26 & 0.32 {\hskip 0.05in}\\ 
Sc II  &  5  & 3.15 & $-$3.09 & 0.25 & 0.28 {\hskip 0.05in}\\ 
Ti II  &  9  & 4.95 & $-$3.07 & 0.27 & 0.28 {\hskip 0.05in}\\ 
Cr I  &  3  & 5.64 & $-$3.46 & $-$0.12 & 0.28 {\hskip 0.05in}\\ 
Mn I  &  3  & 5.43 & $-$3.81 & $-$0.47 & 0.28 {\hskip 0.05in}\\ 
Fe I  &  46  & 7.50 & $-$3.34 & 0.00 & 0.28 {\hskip 0.05in}\\ 
Fe II  &  3  & 7.50 & $-$3.33 & 0.01 & 0.34 {\hskip 0.05in}\\ 
Sr II  &  2  & 2.87 & $-$4.64 & $-$1.30 & 0.28 {\hskip 0.05in}\\ 
Ba II  &  1  & 2.18 & $-$4.19 & $-$0.85 & 0.28 {\hskip 0.05in}\\ 
Eu I  &  1  & 0.52 & $<-$2.04 & $<$1.30 & \nodata {\hskip 0.05in}\\ 
\enddata
\tablenotetext{a}{Random uncertainties. See Table~\ref{tab:uncertainties} for total uncertainties.}
\tablenotetext{b}{Corrected for the evolutionary status of the star following \citet{pfb+14}.}
\tablenotetext{c}{Colons (:) indicate large uncertainties despite the detection of a line feature.}
\end{deluxetable*}

\startlongtable
\begin{deluxetable*}{lrrrr|lrrrr} 
\tabletypesize{\footnotesize}
\tablecolumns{10}
\tablecaption{Uncertainties\label{tab:uncertainties}}
\tablehead{   
 \colhead{El.} & 
 \colhead{N} &
  \colhead{$\sigma_{\text{rand}}$} &
  \colhead{$\sigma_{\text{sys}}$} & 
  \colhead{$\sigma_{\text{tot}}$} & 
  \colhead{El.} &
  \colhead{N} &
  \colhead{$\sigma_{\text{rand}}$} &
  \colhead{$\sigma_{\text{sys}}$} & 
  \colhead{$\sigma_{\text{tot}}$} 
}
\startdata
\multicolumn{5}{c}{\textbf{TucII-006}} &
\multicolumn{5}{c}{\textbf{TucII-011}}\\
\bottomrule\\
CH & 2 & 0.27 & 0.51 & 0.58 {\hskip 0.05in} & {\hskip 0.05in} CH & 2 & 0.31 & 0.39 & 0.50 \\ 
Na I & 2 & 0.27 & 0.39 & 0.47 {\hskip 0.05in} & {\hskip 0.05in} Na I & 2 & 0.23 & 0.33 & 0.40 \\ 
Mg I & 2 & 0.37 & 0.32 & 0.49 {\hskip 0.05in} & {\hskip 0.05in} Mg I & 5 & 0.16 & 0.24 & 0.29 \\ 
Al I & 1 & \nodata & \nodata & \nodata {\hskip 0.05in} & {\hskip 0.05in} Al I & 2 & \nodata & \nodata & \nodata \\ 
Si I & 1 & \nodata & \nodata & \nodata {\hskip 0.05in} & {\hskip 0.05in} Si I & 2 & \nodata & \nodata & \nodata \\ 
Ca I & 2 & 0.27 & 0.30 & 0.40 {\hskip 0.05in} & {\hskip 0.05in} Ca I & 10 & 0.21 & 0.16 & 0.26 \\ 
Sc II & 5 & 0.27 & 0.28 & 0.39 {\hskip 0.05in} & {\hskip 0.05in} Sc II & 6 & 0.16 & 0.15 & 0.22 \\ 
Ti II & 12 & 0.27 & 0.64 & 0.69 {\hskip 0.05in} & {\hskip 0.05in} Ti I & 2 & 0.16 & 0.20 & 0.26 \\ 
Cr I & 3 & 0.27 & 0.53 & 0.59 {\hskip 0.05in} & {\hskip 0.05in} Ti II & 17 & 0.22 & 0.23 & 0.32 \\ 
Mn I & 3 & 0.27 & 0.29 & 0.40 {\hskip 0.05in} & {\hskip 0.05in} Cr I & 5 & 0.20 & 0.29 & 0.35 \\ 
Fe I & 52 & 0.27 & 0.28 & 0.39 {\hskip 0.05in} & {\hskip 0.05in} Mn I & 3 & 0.17 & 0.22 & 0.28 \\ 
Fe II & 3 & 0.27 & 0.20 & 0.34 {\hskip 0.05in} & {\hskip 0.05in} Fe I & 99 & 0.16 & 0.23 & 0.28 \\ 
Co I & 1 & 0.34 & 0.57 & 0.67 {\hskip 0.05in} & {\hskip 0.05in} Fe II & 11 & 0.16 & 0.20 & 0.26 \\ 
Ni I & 2 & 0.66 & 0.45 & 0.80 {\hskip 0.05in} & {\hskip 0.05in} Co I & 3 & 0.21 & 0.25 & 0.33 \\ 
Sr II & 2 & 0.27 & 0.22 & 0.35 {\hskip 0.05in} & {\hskip 0.05in} Ni I & 3 & 0.22 & 0.29 & 0.36 \\ 
Ba II & 2 & 0.27 & 0.26 & 0.37 {\hskip 0.05in} & {\hskip 0.05in} Sr II & 2 & 0.19 & 0.21 & 0.28 \\ 
Eu I & 1 & \nodata & \nodata & \nodata {\hskip 0.05in} & {\hskip 0.05in} Ba II & 1 & 0.20 & 0.20 & 0.28 \\ 
& & & & {\hskip 0.05in} & {\hskip 0.05in} Eu I & 1 & \nodata & \nodata & \nodata \\ 
\bottomrule\\
\multicolumn{5}{c}{\textbf{TucII-033}} &
\multicolumn{5}{c}{\textbf{TucII-052}}\\
\bottomrule\\
CH & 2 & 0.18 & 0.41 & 0.45 {\hskip 0.05in} & {\hskip 0.05in} CH & 2 & 0.20 & 0.43 & 0.47 \\ 
Na I & 2 & 0.12 & 0.27 & 0.30 {\hskip 0.05in} & {\hskip 0.05in} Na I & 2 & 0.29 & 0.27 & 0.40 \\ 
Mg I & 4 & 0.12 & 0.25 & 0.28 {\hskip 0.05in} & {\hskip 0.05in} Mg I & 2 & 0.20 & 0.33 & 0.39 \\ 
Al I & 2 & \nodata & \nodata & \nodata {\hskip 0.05in} & {\hskip 0.05in} Al I & 1 & \nodata & \nodata & \nodata \\ 
Si I & 2 & \nodata & \nodata & \nodata {\hskip 0.05in} & {\hskip 0.05in} Si I & 2 & \nodata & \nodata & \nodata \\ 
Ca I & 8 & 0.12 & 0.14 & 0.18 {\hskip 0.05in} & {\hskip 0.05in} Ca I & 4 & 0.20 & 0.17 & 0.26 \\ 
Sc II & 5 & 0.12 & 0.20 & 0.23 {\hskip 0.05in} & {\hskip 0.05in} Sc II & 5 & 0.20 & 0.16 & 0.26 \\ 
Ti I & 2 & 0.51 & 0.22 & 0.56 {\hskip 0.05in} & {\hskip 0.05in} Ti II & 12 & 0.25 & 0.19 & 0.31 \\ 
Ti II & 12 & 0.27 & 0.21 & 0.34 {\hskip 0.05in} & {\hskip 0.05in} Cr I & 5 & 0.20 & 0.28 & 0.34 \\ 
Cr II & 1 & 0.32 & 0.18 & 0.36 {\hskip 0.05in} & {\hskip 0.05in} Mn I & 4 & 0.21 & 0.22 & 0.30 \\ 
Cr I & 5 & 0.12 & 0.28 & 0.30 {\hskip 0.05in} & {\hskip 0.05in} Fe I & 56 & 0.20 & 0.28 & 0.34 \\ 
Mn I & 3 & 0.30 & 0.31 & 0.43 {\hskip 0.05in} & {\hskip 0.05in} Fe II & 3 & 0.20 & 0.14 & 0.24 \\ 
Fe I & 92 & 0.12 & 0.23 & 0.26 {\hskip 0.05in} & {\hskip 0.05in} Co I & 3 & 0.20 & 0.28 & 0.34 \\ 
Fe II & 12 & 0.14 & 0.18 & 0.23 {\hskip 0.05in} & {\hskip 0.05in} Ni I & 2 & 0.20 & 0.28 & 0.34 \\ 
Co I & 2 & 0.26 & 0.30 & 0.40 {\hskip 0.05in} & {\hskip 0.05in} Sr II & 2 & 0.20 & 0.23 & 0.30 \\ 
Sr II & 2 & 0.26 & 0.19 & 0.32 {\hskip 0.05in} & {\hskip 0.05in} Ba II & 2 & 0.20 & 0.19 & 0.28 \\ 
Ba II & 2 & 0.39 & 0.16 & 0.42 {\hskip 0.05in} & {\hskip 0.05in} Eu I & 1 & 0.20 & 0.16 & 0.26 \\ 
Eu I & 1 & \nodata & \nodata & \nodata {\hskip 0.05in} & {\hskip 0.05in} & & & & \\ 
\bottomrule\\
\multicolumn{5}{c}{\textbf{TucII-078}} &
\multicolumn{5}{c}{\textbf{TucII-203}}\\
\bottomrule\\
CH & 2 & 0.20 & 0.39 & 0.44 {\hskip 0.05in} & {\hskip 0.05in} CH & 2 & 0.20 & 0.43 & 0.47 \\ 
Na I & 2 & 0.20 & 0.21 & 0.29 {\hskip 0.05in} & {\hskip 0.05in} Na I & 2 & 0.26 & 0.24 & 0.35 \\ 
Mg I & 5 & 0.20 & 0.25 & 0.32 {\hskip 0.05in} & {\hskip 0.05in} Mg I & 2 & 0.20 & 0.29 & 0.35 \\ 
Al I & 2 & \nodata & \nodata & \nodata {\hskip 0.05in} & {\hskip 0.05in} Al I & 2 & \nodata & \nodata & \nodata \\ 
Si I & 2 & \nodata & \nodata & \nodata {\hskip 0.05in} & {\hskip 0.05in} Si I & 1 & \nodata & \nodata & \nodata \\ 
Ca I & 8 & 0.20 & 0.15 & 0.25 {\hskip 0.05in} & {\hskip 0.05in} Ca I & 4 & 0.20 & 0.19 & 0.28 \\ 
Sc II & 6 & 0.20 & 0.23 & 0.30 {\hskip 0.05in} & {\hskip 0.05in} Sc II & 5 & 0.20 & 0.20 & 0.28 \\ 
Ti I & 1 & 0.36 & 0.25 & 0.44 {\hskip 0.05in} & {\hskip 0.05in} Ti II & 14 & 0.28 & 0.17 & 0.33 \\ 
Ti II & 20 & 0.25 & 0.25 & 0.35 {\hskip 0.05in} & {\hskip 0.05in} V II & 1 & 0.21 & 0.13 & 0.25 \\ 
Cr I & 5 & 0.26 & 0.28 & 0.38 {\hskip 0.05in} & {\hskip 0.05in} Cr I & 3 & 0.20 & 0.29 & 0.35 \\ 
Mn I & 2 & 0.20 & 0.28 & 0.34 {\hskip 0.05in} & {\hskip 0.05in} Mn I & 3 & 0.20 & 0.28 & 0.34 \\ 
Fe I & 83 & 0.20 & 0.23 & 0.30 {\hskip 0.05in} & {\hskip 0.05in} Fe I & 59 & 0.20 & 0.25 & 0.32 \\ 
Fe II & 9 & 0.20 & 0.25 & 0.32 {\hskip 0.05in} & {\hskip 0.05in} Fe II & 4 & 0.20 & 0.13 & 0.24 \\ 
Co I & 1 & 0.42 & 0.29 & 0.51 {\hskip 0.05in} & {\hskip 0.05in} Co I & 2 & 0.20 & 0.25 & 0.32 \\ 
Ni I & 2 & 0.24 & 0.27 & 0.36 {\hskip 0.05in} & {\hskip 0.05in} Ni I & 3 & 0.20 & 0.27 & 0.34 \\ 
Sr II & 1 & \nodata & \nodata & \nodata {\hskip 0.05in} & {\hskip 0.05in} Sr II & 1 & 0.30 & 0.22 & 0.37 \\ 
Ba II & 1 & \nodata & \nodata & \nodata {\hskip 0.05in} & {\hskip 0.05in} Ba II & 1 & 0.20 & 0.18 & 0.27 \\ 
Eu I & 1 & \nodata & \nodata & \nodata {\hskip 0.05in} & {\hskip 0.05in} Eu I & 1 & \nodata & \nodata & \nodata \\ 
\bottomrule\\
\multicolumn{5}{c}{\textbf{TucII-206}}\\
\bottomrule\\
CH & 2 & 0.28 & 0.51 & 0.58 {\hskip 0.05in}\\ 
Na I & 2 & 0.46 & 0.32 & 0.56 {\hskip 0.05in}\\ 
Mg I & 3 & 0.28 & 0.31 & 0.42 {\hskip 0.05in}\\ 
Al I & 2 & \nodata & \nodata & \nodata {\hskip 0.05in}\\ 
Si I & 1 & \nodata & \nodata & \nodata {\hskip 0.05in}\\ 
Ca I & 2 & 0.32 & 0.27 & 0.42 {\hskip 0.05in}\\ 
Sc II & 5 & 0.28 & 0.26 & 0.38 {\hskip 0.05in}\\ 
Ti II & 9 & 0.28 & 0.29 & 0.40 {\hskip 0.05in}\\ 
Cr I & 3 & 0.28 & 0.29 & 0.40 {\hskip 0.05in}\\ 
Mn I & 3 & 0.28 & 0.24 & 0.37 {\hskip 0.05in}\\ 
Fe I & 46 & 0.28 & 0.27 & 0.39 {\hskip 0.05in}\\ 
Fe II & 3 & 0.34 & 0.29 & 0.45 {\hskip 0.05in}\\ 
Sr II & 2 & 0.28 & 0.25 & 0.38 {\hskip 0.05in}\\ 
Ba II & 1 & 0.28 & 0.29 & 0.40 {\hskip 0.05in}\\ 
Eu I & 1 & \nodata & \nodata & \nodata {\hskip 0.05in}\\ 
\enddata
\end{deluxetable*}

\startlongtable
\begin{deluxetable*}{lllrrrr} 
\tablecolumns{6}
\footnotesize
\tablewidth{\textwidth}
\tablecaption{Line measurements\label{tab:linetable}}
\tablehead{   
  \colhead{Star} &
  \colhead{Rest} &
  \colhead{Species} & 
  \colhead{Excitation} & 
  \colhead{Oscillator} &
  \colhead{Equivalent} & 
  \colhead{$\log\epsilon (\rm{X})$} \\
\colhead{name} &
\colhead{wavelength (\AA)} &
  \colhead{} & 
  \colhead{potential (eV)} & 
  \colhead{strength} &
  \colhead{width (m\AA)} & 
  \colhead{} 
}
\startdata
TucII-006 & 4313.00 & CH & syn & syn & syn & 6.25 \\
TucII-006 & 4323.00 & CH & syn & syn & syn & 6.50 \\
TucII-006 & 5889.95 & Na I & 0.00 & 0.11 & 146.4 & 3.61 \\
TucII-006 & 5895.92 & Na I & 0.00 & $-$0.19 & 114.9 & 3.31 \\
TucII-006 & 5172.68 & Mg I & 2.71 & $-$0.45 & 120.3 & 4.33 \\
TucII-006 & 5183.60 & Mg I & 2.72 & $-$0.24 & 162.6 & 4.92 \\
TucII-006 & 3944.00 & Al I & syn & syn & syn & 2.83 \\
TucII-006 & 3905.52 & Si I & syn & syn & syn & 4.80 \\
TucII-006 & 4454.78 & Ca I & 1.90 & 0.26 & 73.2 & 3.77 \\
TucII-006 & 6122.22 & Ca I & 1.89 & $-$0.32 & 42.3 & 3.69 \\
\enddata
\tablecomments{Table~\ref{tab:linetable} is published in its entirety in the machine-readable format.
      A portion is shown here for guidance regarding its form and content.}
\end{deluxetable*}

\end{document}